%% file: slac-pub-9735.tex
\documentclass[12pt]{article}

 \usepackage{graphicx}
\usepackage{rotating}
\begin{document}

\def\int{\ifmmode\mathrm{I}
          \else$\mathrm{I}$\fi}
\def\refl{\ifmmode\mathrm{\mathcal R}
          \else{$\mathcal R$}\fi}

\def\babar{\mbox{\sl B\hspace{-0.4em} {\small\sl A}\hspace{-0.37em} \sl B\hspace{-0.4em} {\small\sl A\hspace{-0.02em}R}}}
\def\dirc       {{\sc DIRC}}

\thispagestyle{empty}
\renewcommand{\thefootnote}{\fnsymbol{footnote}}

\begin{flushright}
{\small
SLAC--PUB--9735\\
April 2003\\}
\end{flushright}

\vspace{.8cm}

\begin{center}
{\bf\large   
Optical Properties of the DIRC Fused Silica Cherenkov 
Radiator\footnote{Work supported by
Department of Energy contract  DE--AC03--76SF00515.}}

\vspace{1cm}

J. Cohen-Tanugi\footnote{Present address: Universit\`a di Pisa, Scuola Normale
  Superiore and INFN, I-56010 Pisa, Italy.}, 
M. Convery,
B. Ratcliff,
X. Sarazin
\footnote{Present address: Laboratoire de l'Acc\'el\'erateur Lin\'eaire, F-91898 Orsay, France.},
J. Schwiening 
and
J. Va'vra\\
\vskip 0.1in
Stanford Linear Accelerator Center, Stanford University, 
Stanford, CA 94309, USA

\end{center}

\vfill

\begin{center}
{\bf\large   
Abstract }
\end{center}

\begin{quote}
The \dirc\
is a new type of Cherenkov detector that is successfully operating 
as the hadronic particle identification system for the \babar\ experiment 
at SLAC. 
The fused silica bars that serve as the \dirc's Cherenkov radiators 
must transmit the light over long 
optical pathlengths with a large number of internal reflections.
This imposes a number of stringent and novel requirements on
the bar properties.
This note summarizes a large amount of R\&D that was performed
both to develop specifications and production methods and 
to determine whether commercially produced bars could meet the requirements.
One of the major outcomes of this R\&D work is an understanding of
methods to select radiation hard and optically uniform fused silica
material. 
Others include measurement of the wavelength dependency of the
internal reflection coefficient, 
and its sensitivity to surface contaminants, 
development of radiator support methods, 
and selection of good optical glue.
\end{quote}

\vfill

\begin{center} 
{\it Submitted to Nuclear Instruments and Methods A} 
\end{center}

\newpage



%
\pagestyle{plain}

\input body.tex

\end{document}

%% file: body.tex
\input intro.tex
\input basic.tex

\input transrefl.tex

\input radiation.tex

\input shims.tex

\input mech.tex

\input concl.tex
\input biblio.tex

%% file: intro.tex
\section{Introduction}
\label{sec:intro}

The Detector of Internally Reflected Cherenkov light (\dirc)~\cite{1} is a
new type of Cherenkov ring imaging detector that  has been operating
successfully at the \babar\ experiment at SLAC~\cite{babarnim} for
over three years. 
The device uses synthetic fused silica bars (colloquially called
quartz bars), which serve both as the Cherenkov radiator and as light
guides transmitting the photons to an array of ~11,000 photomultiplier
tubes (PMT). 

\begin{figure}
\centerline{\includegraphics[width=11cm]{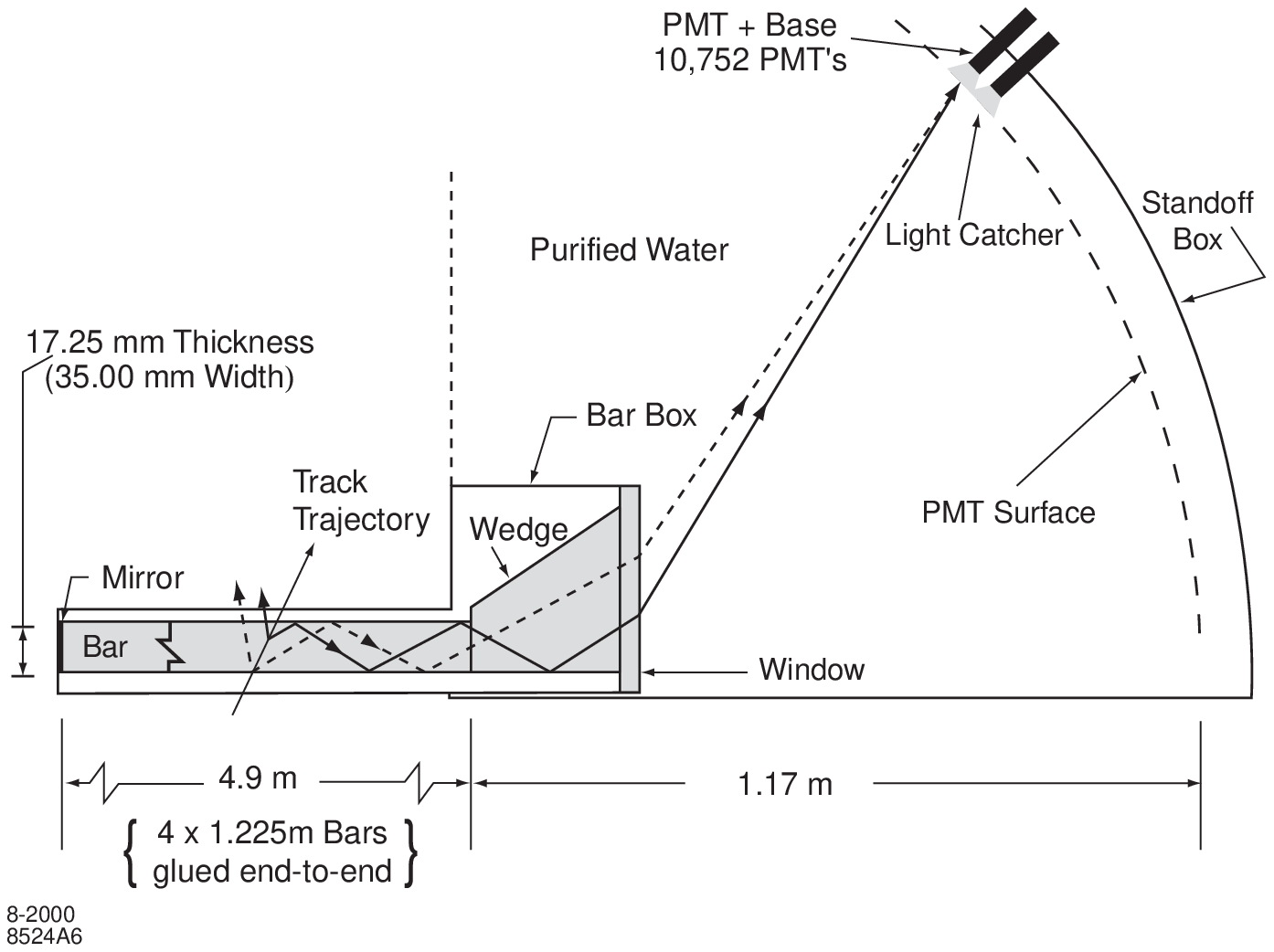}}
\caption{		
\label{Fig:1}
Imaging of Cherenkov photons with \dirc\ fused silica bar. 
}
\end{figure}

Figure~\ref{Fig:1} illustrates the principle of the \dirc.
A fraction of the Cherenkov photons produced by tracks
passing through the bars is trapped by total internal reflection and
propagates down the bars with very little loss and with the Cherenkov
angle preserved (up to reflection ambiguities).
A mirror at the far end reflects those photons that were originally 
traveling away from the detection end of the bar. 
At the detector end of the bar, photons pass through a fused silica wedge
that allows photons to begin to expand into the detector region,
while reflecting those photons that would otherwise
miss the PMTs back into the photon detectors.
Photons then pass through a window that separates the box holding 
the bars from the detection region.
Water is the optical coupling medium between the bars and the PMT detector. 
The distance between the bar box window and the PMT detector is about
1.1\ m. 

\begin{figure}
\centerline{\includegraphics[width=11cm]{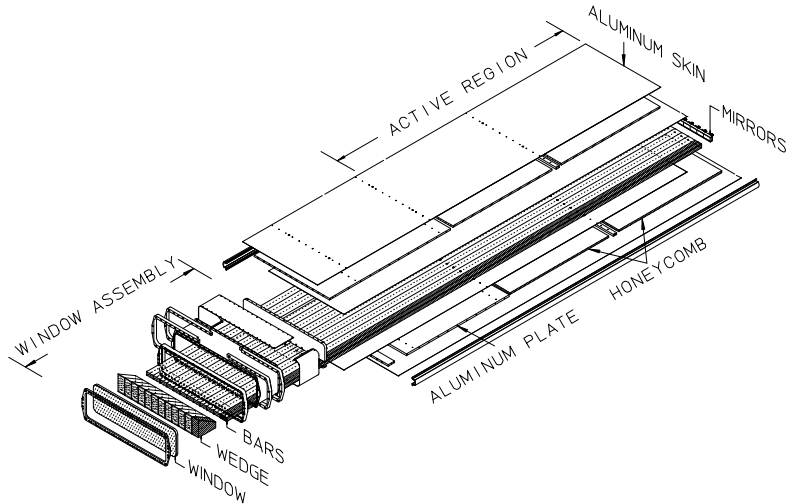}}
\caption{		
\label{Fig:2}
\dirc\ fused silica Cherenkov radiator box as implemented in the
\babar\ experiment at SLAC. Twelve 4.9\ m long fused silica bars are
located in each bar box.
}
\end{figure}

\begin{figure}
\centerline{{\includegraphics[width=11cm]{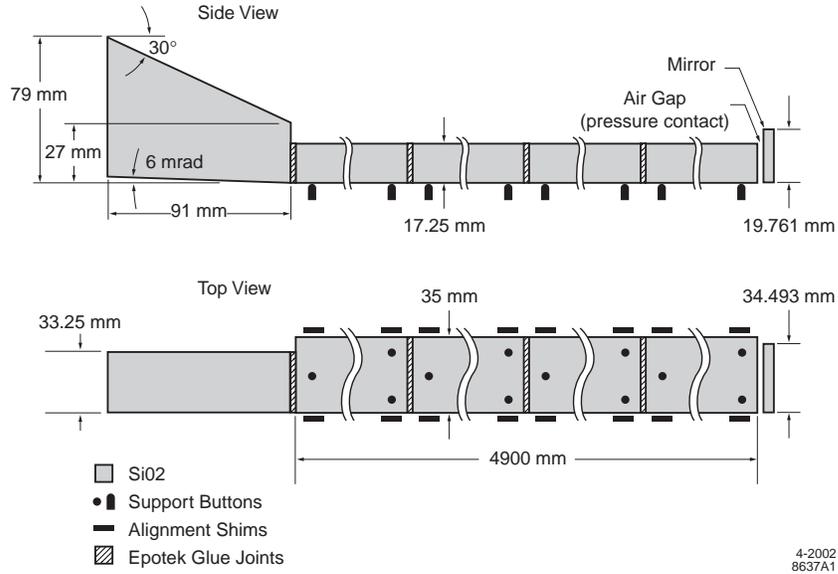}}}
\caption{Schematics of a single \dirc\ radiator bar, oriented as shown
  in \protect{Figure~\ref{Fig:2}}, in side and top view.} \label{fig:radiator}
\end{figure}

Figure~\ref{Fig:2} shows a single bar box with 12 bars. 
There are 12 such boxes in the \babar\ detector.
Each bar, schematically shown in Figure~\ref{fig:radiator}, 
measures 4.9\ m in length and is made 
of four short segments glued together with Epotek 301-2 optical 
epoxy~\cite{epotek}. 
Each short segment has the following design specifications: 
length 1225\ mm, width 35.0\ mm, and thickness 17.2\ mm. 
The 1225$\times$35.0 mm surface of the bar is called the ``face'',
the 1225$\times$17.2 mm surface is the ``side'', and the 
35.0$\times$17.2 mm surface is the ``end''.
To preserve the image resolution properties, the original specification 
was that any bar face-to-side angle be within 
$\pm$ 0.25\ mrad of 90$^{\circ}$ 
(see section~\ref{sec:mech} for the final compromise). 
The bars must have sharp edges without too many chips to
limit photon losses (specifications required less than 6\ mm$^2$ of
damaged surface per bar). 
To check that the mechanical specifications are within our tolerances, 
we built a digital microscope system with 
image reconstruction software. 

Since the trapped Cherenkov photons must typically bounce several 
hundred times before they exit the bar, the internal reflection 
coefficient must be close to one -- less than 0.999 is
unacceptable. 
Losses as small as 10$^{-4}$ per bounce must be
measured during the quality assurance process. 
Since such small losses are difficult to measure in a single
bounce, we developed a method to measure the reflection
coefficient by combining $\sim$50 bounces in a single measurement. 
The high surface reflectivity must be maintained for 
the expected ten-year lifetime of the experiment. 
Therefore, we studied the possible
influence of contamination of the
fused silica surfaces on the reflectivity
coefficient. 

The bulk transmission of the fused silica must also be very good.
Attenuation lengths must be $\ge$ 100 meters.
This high transmission must be maintained after exposure to
high levels of ionizing radiation.
This is because the \babar\ detector operates at a high
luminosity $e^+e^-$ machine that has a considerable
gamma ray background.
The optical glues must be similarly radiation hard.
The luminosity of PEP II is currently
$\sim$5$\cdot 10^{33}$ cm$^{-2}$sec$^{-1}$
and should exceed 10$^{34}$ cm$^{-2}$sec$^{-1}$ in one to two years.

%% file: basic.tex
\section{Basic Properties of Fused Silica}
\label{sec:material}
\subsection{Types of Fused Silica}
The crystalline form of Silicon Dioxide (SiO$_2$) is called quartz and is
the second most abundant mineral on earth. 
Quartz crystals are birefringent and are not suitable for use in the DIRC.
An amorphous form of SiO$_2$, which we will call
natural fused silica, may be formed by crushing and melting natural quartz
in a hydrogen flame.
Optical grades of 
this material possess many of the properties required by the DIRC, such
as long transmission length and good polishability.
However, it suffers from the high level of impurities typically found in the
natural quartz used to produce it, which can lead to 
significant sensitivity to ionizing radiation.
A third material, which we will call synthetic fused silica, has been 
commercially produced for the past few decades.
It is formed artificially by burning silicon
tetra-chloride (SiCl$_4$), or other feedstock in an oxygen atmosphere.
This material can be made very pure and is widely used in the
fiber optics industry. 
All of these forms of SiO$_2$ are colloquially called quartz, but, 
strictly speaking, this term only applies to the crystalline form.

Both natural fused silica and synthetic fused silica may be obtained
from a number of manufacturers.
We have tested a number of brands of 
natural fused silica, including Vitreosil-F~\cite{qpc2}, and a number
of brands of synthetic fused silica, including Suprasil~\cite{heraeus}, 
Spectrosil 2000 and Spectrosil B~\cite{qpc}.

\begin{figure} 
\centerline{\includegraphics[width=11cm]{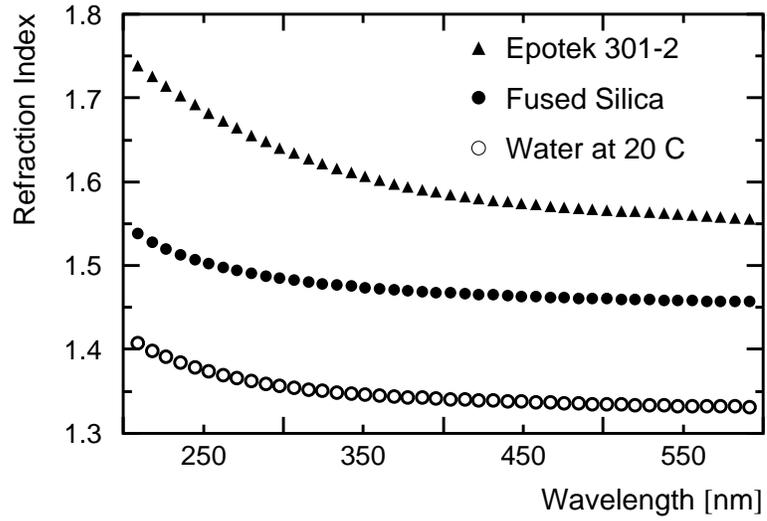}}
\caption{		
\label{Fig:3}
Wavelength dependence of the refraction indices for fused
silica, water, and Epotek 301-2 optical glue.  
The DIRC acceptance bandwidth starts at 300\ nm~\cite{4}.}
\end{figure}
	
\begin{figure} 
\centerline{\includegraphics[width=11cm]{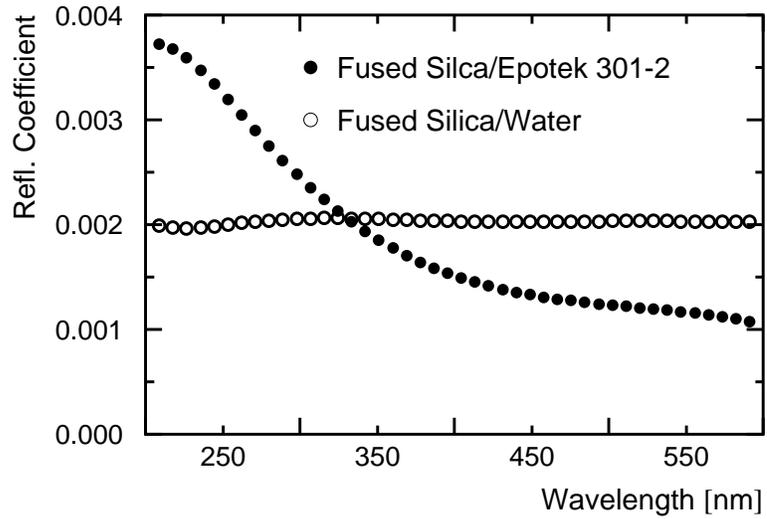}}
\caption{		
\label{Fig:4}
Wavelength dependence of the calculated Fresnel
reflection coefficients for a
single fused silica-water, and a fused silica-Epotek 301-2 interface 
at an incidence angle of 0$^{\circ}$~\cite{4}.
}\end{figure}

\subsection{Refractive Index}
\label{sec:refr}
To account for various effects in the DIRC, it is necessary to know
the refractive index of the optical components of the DIRC.
Figure~\ref{Fig:3} shows the refractive index for
fused silica~\cite{2}, water~\cite{3} and Epotek 301-2~\cite{4}, the
optical glue used for joining the bars and other optical components.
Figure~\ref{Fig:4} shows the calculated Fresnel
reflection coefficients at an incidence angle of 0$^{\circ}$ for the boundaries of
fused silica/water and fused silica/Epotek 301-2. 
Due to a good
match of the fused silica and glue refraction indices, the reflection
coefficient is small (less than 0.5\%) and almost independent of 
wavelength above 300\ nm ~\cite{5}.

\begin{figure} 
\vspace*{-3mm}
\centerline{\includegraphics[width=11cm]{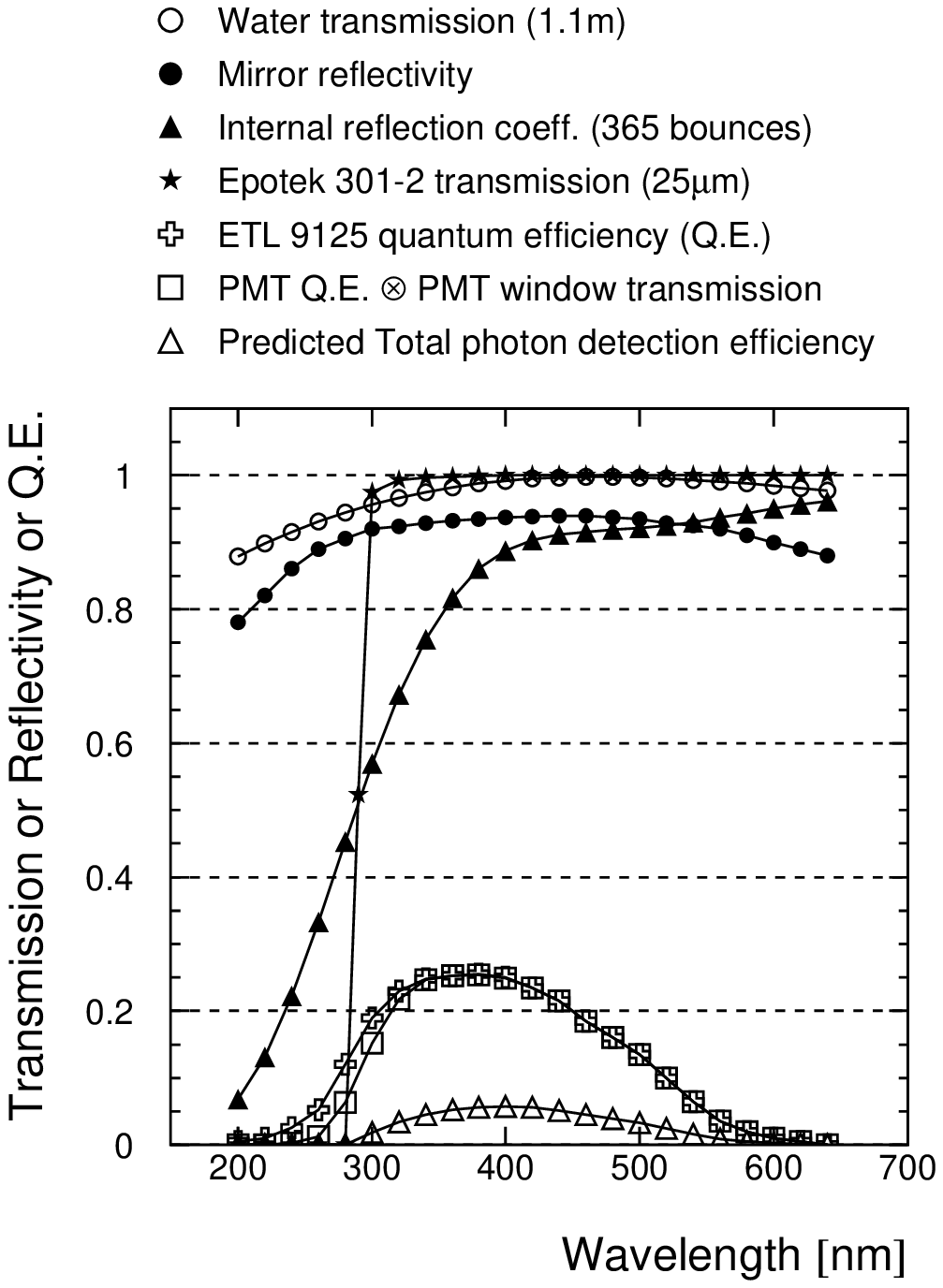}}
\caption{		
\label{Fig:5}
Contributions to the photoelectron detection efficiency for
a track perpendicular to the fused silica bar located in the middle of
the \babar\ sensitive area.
The DIRC final efficiency takes into account many other correction factors as
well~\cite{5}.
}\end{figure}

\subsection{Optical Bandwidth of DIRC}
\label{sec:dirceffi}
Figure~\ref{Fig:5} shows the optical bandwidth of the DIRC 
detector for a typical Cherenkov photon emitted by a particle at
90$^{\circ}$ polar angle in \babar~\cite{5}. 
Contributions to the bandwidth include: 
bar and water transmission, 
mirror reflectivity, 
internal reflection coefficient (assuming 365 bounces), 
glue (Epotek 301-2) transmission and 
PMT quantum efficiency (ETL 9125FLB17 PMT~\cite{emi}) and window transmission.
One can see that the bandwidth is cut at $\sim$300\ nm, 
mainly due to the Epotek 301-2 glue transmission properties. 
This is actually advantageous for the
DIRC because it reduces the chromatic error contributions to the
overall Cherenkov angle and timing resolution. 
In addition, it makes the DIRC sensitive mainly in the visible 
and near-visible UV
wavelength regions adding many practical advantages from a 
construction point of view.

\subsection{Optical Homogeneity of the Synthetic Fused Silica Material}
\label{sec:lobes}
\begin{figure} 
\centerline{\includegraphics[width=11cm]{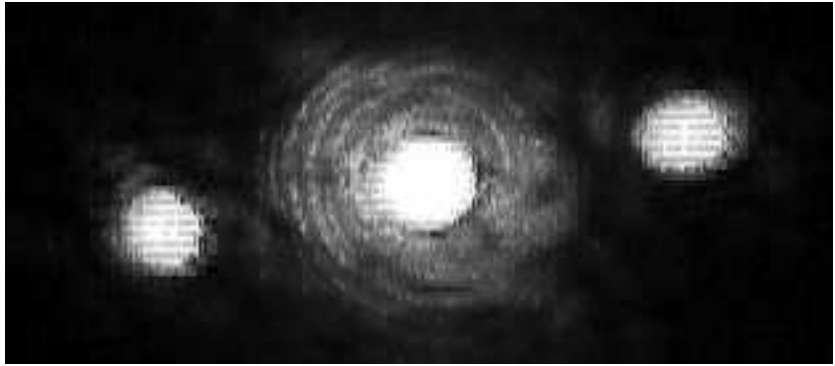}}
\caption{               
\label{Fig:6}
A lobe pattern created with a HeNe laser in the synthetic
fused silica Suprasil Standard~\cite{6}.
}
\end{figure}

One unexpected feature of the synthetic fused silica
used in the DIRC is that it
possesses an observable periodic optical non-homogeneity in its
volume~\cite{6}. 
This effect was initially discovered while shining a
laser beam at non-zero angles relative to the bar axis and observing
that some light was scattered into a diffraction-like pattern, called
``lobes'' (see Figure~\ref{Fig:6}). 
This effect is observed in both types of synthetic fused
silica considered for use in the DIRC:  
Heraeus Suprasil and QPC Spectrosil.

\begin{figure} 
\centerline{\includegraphics[width=11cm]{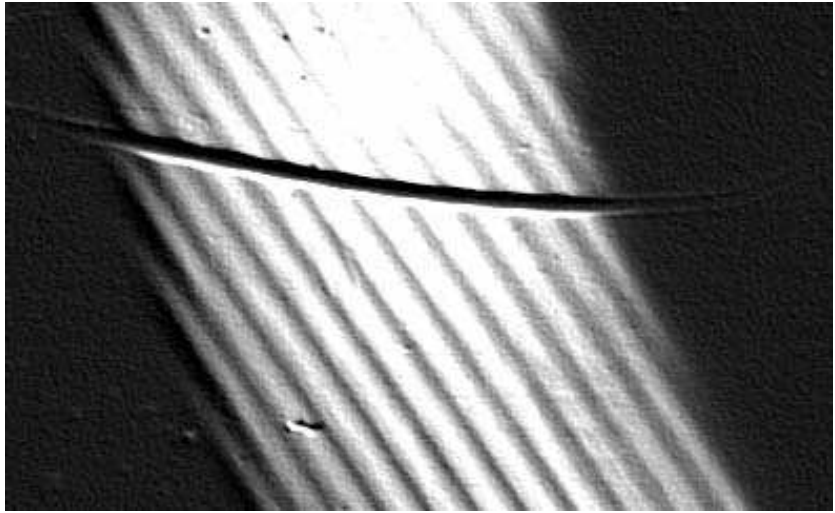}}
\caption{               
\label{Fig:7}
Image observed under a microscope using a ``white''
ring-type light. The periodic structure is evident. The dark line
across the top is a 100 $\mu$m wire placed on the top surface and used for
scaling. No obvious periodic pattern was observed with 
Spectrosil 2000~\cite{6}.
}
\end{figure}

\begin{figure} 
\centerline{\includegraphics[width=11cm]{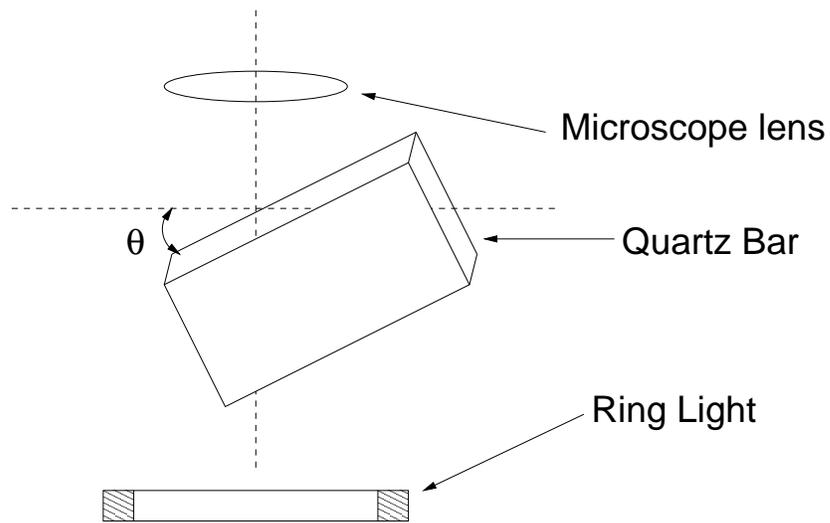}}
\caption{		
\label{Fig:8}
A microscope setup to observe the ring pattern as shown in
Figure~\ref{Fig:7}~\cite{6}.
}\end{figure}

This ``lobe effect'' must be taken very seriously because it can
cause photon losses or image distortion of the Cherenkov light. 
A few additional features of the lobe effect are:
\begin{itemize}
\item{The lobes are produced over a range of incident 
angles typically greater than 45 degrees 
(or less than 135$^{\circ}$) relative to the bar axis.}
\item{Bars that produce lobes for incident angles less than 90
degrees relative to the bar axis do not produce lobes for 
angles greater than 90 degrees, and vice-versa.}
\item{Some bars produced no lobes at all.}
\end{itemize}
The lobe pattern is suggestive of diffraction from
a periodic structure in the fused silica~\cite{9}. 
Furthermore, a clear periodic pattern 
could also be directly observed using ordinary white light
(Figure~\ref{Fig:7}) with an optical digital
microscope configured as shown in Figure~\ref{Fig:8}. 
The same ``directionality'' was observed as for lobe 
production - {\it i.e.} if the pattern was observed
for $\theta>0$ it was not for $\theta<0$. 
Also, the pattern persisted after removing ~2\ mm of material from 
each surface of the bar, thereby establishing that the pattern is 
in the volume of the fused silica material.

\begin{figure} 
\centerline{\includegraphics[width=11cm]{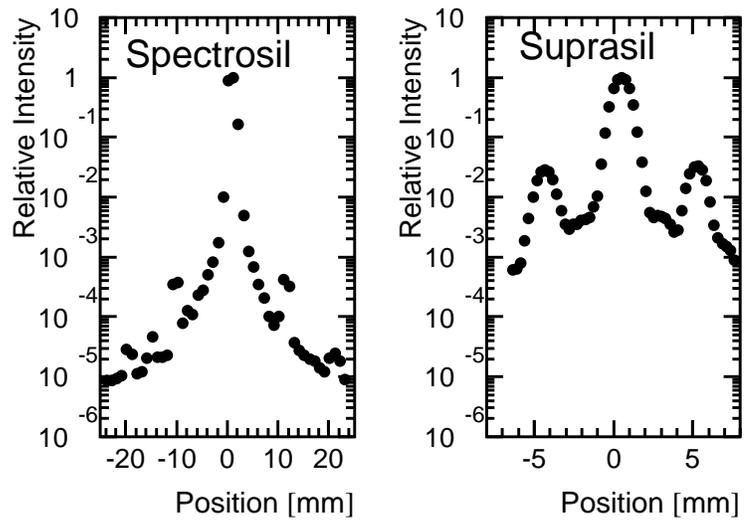}}
\caption{		
\label{Fig:9}
The relative lobe power observed using a HeNe laser (633\ nm)
for QPC Spectrosil fused silica (left) and Heraeus Suprasil 
material (right). Note different scales~\cite{6}.
}\end{figure}

Although the lobe effect was initially observed in Heraeus Suprasil fused
silica, it is also present, although at lower intensity and at larger
angles relative to the bar axis, in QPC Spectrosil.
Figure~\ref{Fig:9} shows the output of a photodiode as it is
scanned across the lobe patterns of the two types of fused silica
measuring the relative intensity. 
Evidently, the intensity of the
lobes from QPC Spectrosil fused silica material ($\sim3 \cdot 10 ^{-4}$)
is much smaller than
from Heraeus Suprasil ($\sim3\cdot 10^{-2}$).  

\begin{figure} 
\centerline{\includegraphics[width=11cm]{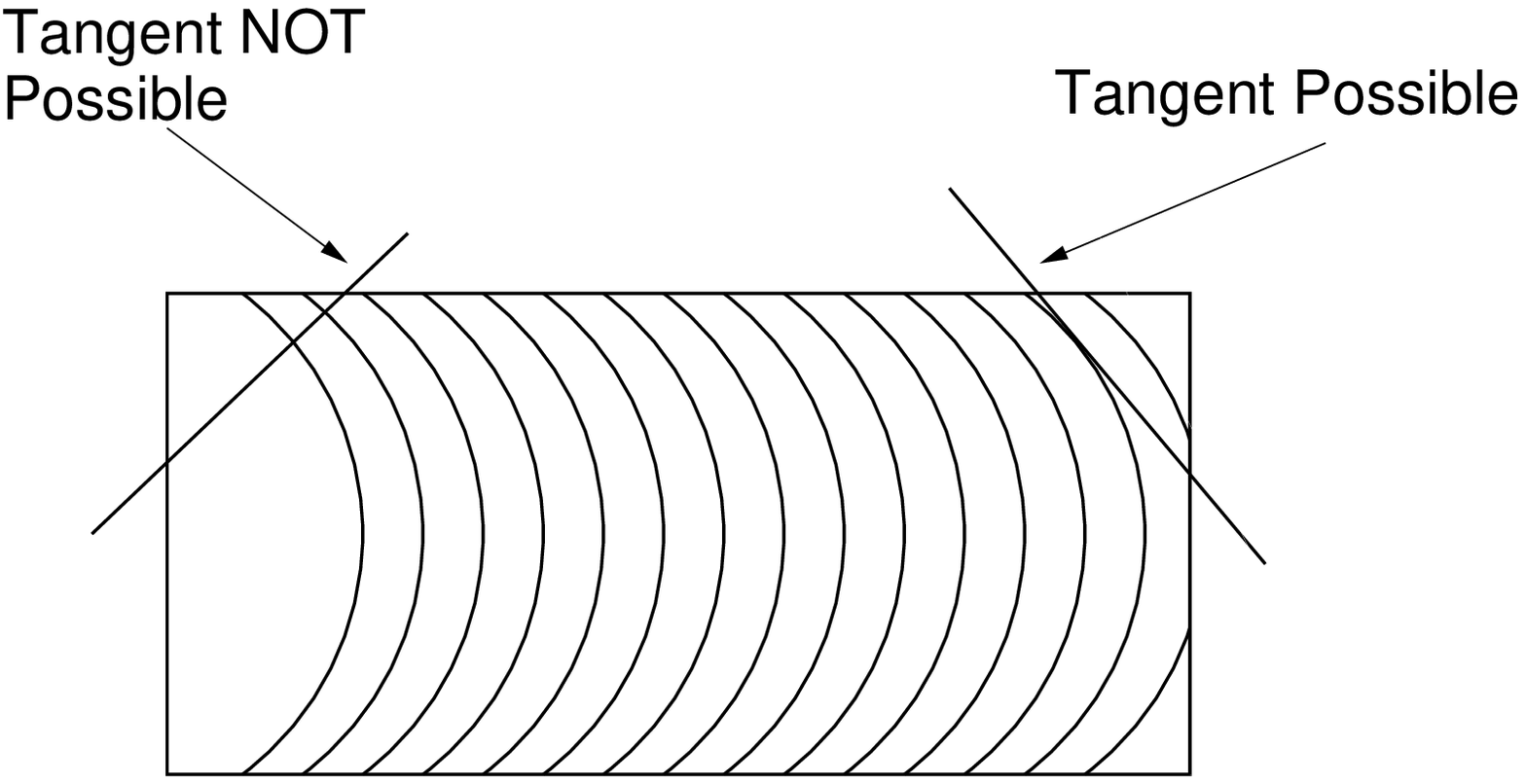}}
\caption{		
\label{Fig:10}
A possible explanation of the origin of the interference is in
the layering of the SiO$_2$ material within the ingot.
}\end{figure}

Figure~\ref{Fig:10} shows a model of the structure of
the optical inhomogeneity, where it is assumed that there are curved
``layers'' of varying index of refraction within the fused silica ingots
from which the bars are produced. 
If a laser beam is traveling tangent
to these layers, then it would, in effect, see a ``diffraction grating''
formed by the alternating layers of high and low refraction index,
thereby producing lobes. 
If however, the beam is traveling
perpendicular to the layers, no lobes would be produced. This explains
the ``directionality'' described above. 
Furthermore, we would expect
that the opening angle of the lobes would be given by 
$\alpha = \lambda / \mu$, where
$\lambda$  is the wavelength of light and $\mu$ is the spacing of the
layers. 
Also, the phase change produced by the inhomogeneity would be
$\Delta \phi = 4 \pi \mathrm{a} \sqrt{R \mu} / \lambda$, 
where $a$ is the amplitude of the
inhomogeneity ($a = \Delta n /n$)
and R is the radius of curvature of the layers, 
which we assume are circular. 
The fraction of the power in the lobes ($\mathrm{f}$) 
is proportional to the square of the phase change,
$ \mathrm{f} \propto 1 / \lambda^2$. 

\begin{figure} 
\centerline{\includegraphics[width=11cm]{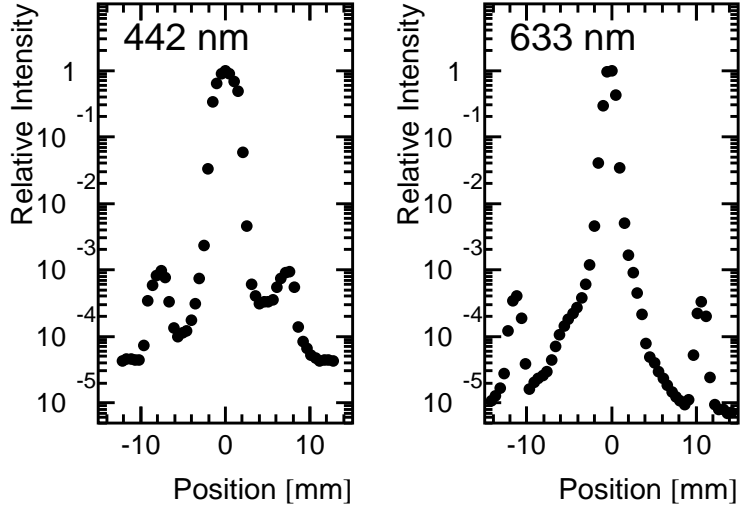}}
\caption{		
\label{Fig:11}
The lobe pattern observed at two different
wavelengths for the QPC Spectrosil material. 
The difference in the width of the
peaks is due to the different spot size of the HeNe and HeCd laser
beams~\cite{6}.
}\end{figure}

Measuring the lobes with
two different light wavelengths can test the model. 
Figure~\ref{Fig:11} shows
the results of the scans with a HeCd laser (442\ nm) and a HeNe laser
(633\ nm). 
Table~\ref{tab:lobewv} shows the ratio of opening angles and lobe
powers predicted by the model to those observed. 
The good agreement
between the model and the measurements gives us confidence that the
basic features of the model are correct. 
Converting the measured power in the lobes to
an inhomogeneity amplitude, a, we find \mbox{a = 6$\cdot 10^{-6}$} 
for Heraeus Suprasil and 1$\cdot 10^{-6}$ for QPC Spectrosil. 
More detailed measurements can also determine the shape of the
layers in the fused silica.

\begin{table}
\begin{center}
\begin{tabular}{l|cc}
& Predicted & Measured \\
\hline
$\alpha_{\rm blue} / \alpha_{ \rm red}$ & 0.70 & 0.69 $\pm$ 0.03 \\
$ f_{\rm blue} / f_{\rm red}$          & 2.05 & 2.58 $\pm$ 0.40 \\
\end{tabular}
\end{center}
\caption{\label{tab:lobewv} Comparison of predicted and measured values
of lobe quantities.}
\end{table}

This model is also consistent with the direct observation of 
structure shown in Figure~\ref{Fig:7}.
This observation could be explained as a modulation of the 
refraction angle of light rays as they enter the fused 
silica. 

Because of its bright lobes and the angular range over which they 
are produced, Heraeus Suprasil was rejected for use in
the DIRC. 
QPC Spectrosil fused silica was deemed acceptable both because of its lower
lobe power and also since the lobes in the QPC fused silica are only 
produced at 
angles close to perpendicular to the bar axis, which are not relevant 
for the photons detected in the DIRC. 
In practice, many QPC Spectrosil ingots have no discernible lobes
at all.
Based on these observations of optical inhomogeneity, 
QPC Spectrosil 2000 was chosen as the material for the DIRC bars.

It is interesting to note that,
at our level of sensitivity, 
no such lobes have been observed in any of
our samples of natural fused silica, nor in the synthetic fused silica
windows used for the CRID TPC detectors and liquid radiators~\cite{7}. 
For all of these materials, the SiO$_2$ material is 
deposited onto a stationary template to form a ``boule''.
What makes the DIRC's synthetic material special
is that it is produced in a long ingot form, where the SiO$_2$ material
is deposited layer-by-layer while the ingot is rotating. 
The details
of this process are proprietary, and so not directly accessible to us,
but it seems plausible that inhomogeneity amplitudes of a few
times $10^{-6}$ could easily be produced by it. 
We should also point out that in most applications of synthetic 
fused silica, such as windows
or optical fibers, the light is expected to travel only at angles
close to parallel with the ``bar axis''. 
The DIRC may be one of the few
applications where light must propagate at large angles with respect
to this axis, and thus be sensitive to the ``lobe effect''.

%% file: transrefl.tex
\section{Transmission and Internal Reflection Coefficient 
of Fused Silica Bars}
\label{sec:transrefl}

For the measurement of the transmission and the coefficient of total
internal reflection \refl\ of the DIRC bars we use two methods. 
A manual method is able to measure the absolute value of \refl\ while
an automated setup, used primarily for the quality assurance measurements
during DIRC bar production, determines the transmission as well 
as the relative \refl\ difference between a reference bar and the
tested bar.

\subsection{Measurement Technique}

\begin{figure} 
\centerline{\includegraphics[width=11cm]{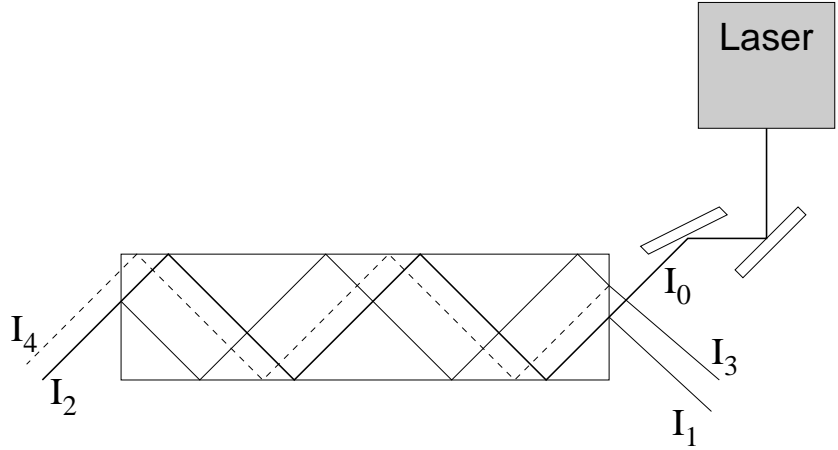}}
\caption{		
\label{Fig:15}
A schematic layout of the setup to measure the
internal reflection coefficient in studies of wavelength 
dependence and tests of surface contamination.
}\end{figure}

The internal reflection coefficient is measured absolutely with 
a ``calorimetric'' method~\cite{13}.
As shown in Figure~\ref{Fig:15}, this method measures 
five light intensities ($I_0, I_1, I_2, I_3$ and $I_4$)
and uses the number of light bounces, the bar dimensions
and the bulk attenuation as inputs.

The fraction of the light transmitted as it propagates down the bar one
time ($1/x$), may be written as a combination of reflection losses and
bulk attenuation:
\begin{equation}
\label{Eqn:x}
x^{-1} = \refl^{N} \cdot 
\mathrm{exp}\left( - \frac{L}{\Lambda}\cdot\sqrt{1+\left( 
\frac{bN}{L}\right) ^2}\right) \ ,
\end{equation}
where N is the number of bounces, L is the length of the fused silica
bar, b is the width of the fused silica bar, and $\Lambda$ is the bulk
attenuation length.

We can then relate the five measured intensities as:
\begin{equation}
\label{Eqn:calori1}
((( \int_0 - \int_1 ) \cdot {1 \over x} - \int_2) \cdot {1 \over x}
- \int_3 ) \cdot {1 \over x} = \int_4 \ .
\end{equation}
This can be simplified to:
\begin{equation}
\label{Eqn:calori2}
\int_0 = \int_1 + \int_2 \cdot x + \int_3 \cdot x^2 + \int_4 \cdot x^3 \ .
\end{equation}
The solution, $x$ of this equation can then be used to calculate the 
\refl. 
To do so, we need to know the attenuation length, $\Lambda$, 
which
is not easy to measure at long wavelengths. 
In our calculations, we use the value\footnote{This 
value is obtained by combining 
manual measurements done on a 5 meter long bar with the measurements
on 1.2 meter bars using the automated method described below.} 
$\Lambda=500 \pm 167$m 
measured at 442\ nm, and scale it to other wavelengths by $1/\lambda^4$,
assuming Rayleigh scattering.
This dependence gives $\Lambda \approx 2100$~m at 633\ nm, 
which would be impossible to measure with good accuracy using 
a ~1.2 m long bar.

A further improvement in accuracy is found by adding another term in
equation~\ref{Eqn:calori2}, which can be calculated from the measured
intensities by integrating the infinite sum of higher order
contributions: 
\begin{equation}
\label{Eqn:calori3}
\int_0 = \int_1 + \int_2 \cdot x + \int_3 \cdot x^2 + \int_4 \cdot x^3
\ + \int_{5+} \ .
\end{equation}
where the correction term reduces to: 
\begin{equation}
\label{Eqn:caloricorr}
\int_{5+} = \frac{\int_{3}^{3}}{\int_2 \cdot (\int_2 - \int_3)} \ .
\end{equation}
This correction increases the coefficient typically by $\sim$ 0.0003.

Another small improvement in the accuracy is a correction for
photodiode response due to the finite size of the laser beam. 
As the beam size for each intensity $\int_i$ changes, the 
correction is different for each $\int_i$ component. 
These corrections vary from 1.006 ($\int_0$) to 1.03
($\int_4$).

All intensities are measured by hand with a single
photodiode (Hamamatsu Co.~\cite{hamamatsu} silicon photodiode
S1337-101BQ).
The photodiode is equipped with 
a UV diffuser to improve the uniformity of its response.
However, it is still necessary to move the diode by hand in
order to find a consistent maximum value.
We call this procedure ``peaking''.
We use a second photodiode to monitor the laser intensity. 
The two photodiodes were synchronized during the readout. 
The laser used is a HeCd laser, Liconix~\cite{liconix} 200 series,
model 4214NB. 
It operates at two wavelengths, is equipped with multimode optics,
and provides a vertically polarized beam with a wavelength of 442 or
325\ nm.  
\\

\begin{figure} 
\centerline{\includegraphics[width=11cm]{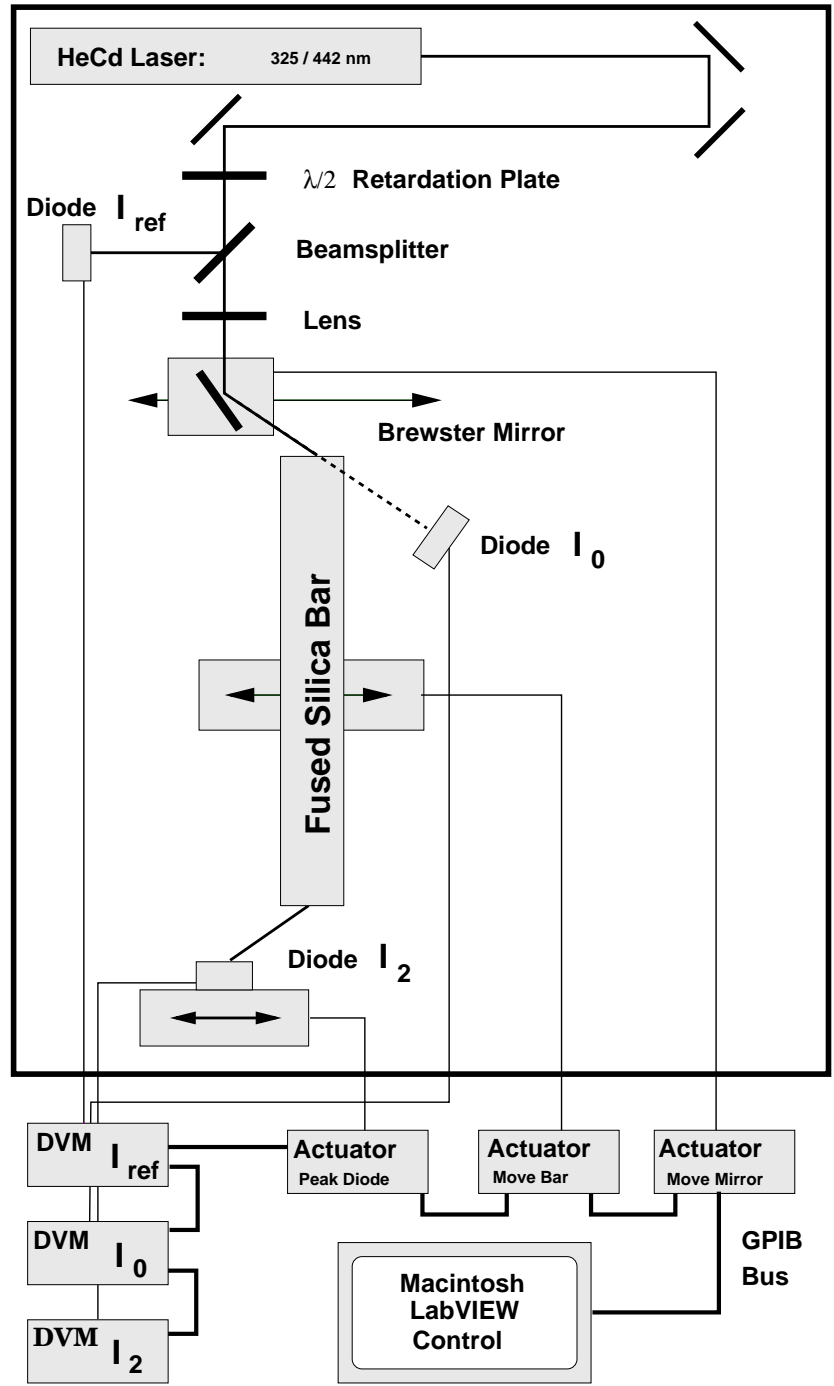}}
\caption{		
\label{Fig:12}
Computer controlled setup used to measure fused silica bar
transmissions and the relative internal reflection coefficient~\cite{10}.
}\end{figure}

\begin{figure} 
\centerline{\includegraphics[width=11cm]{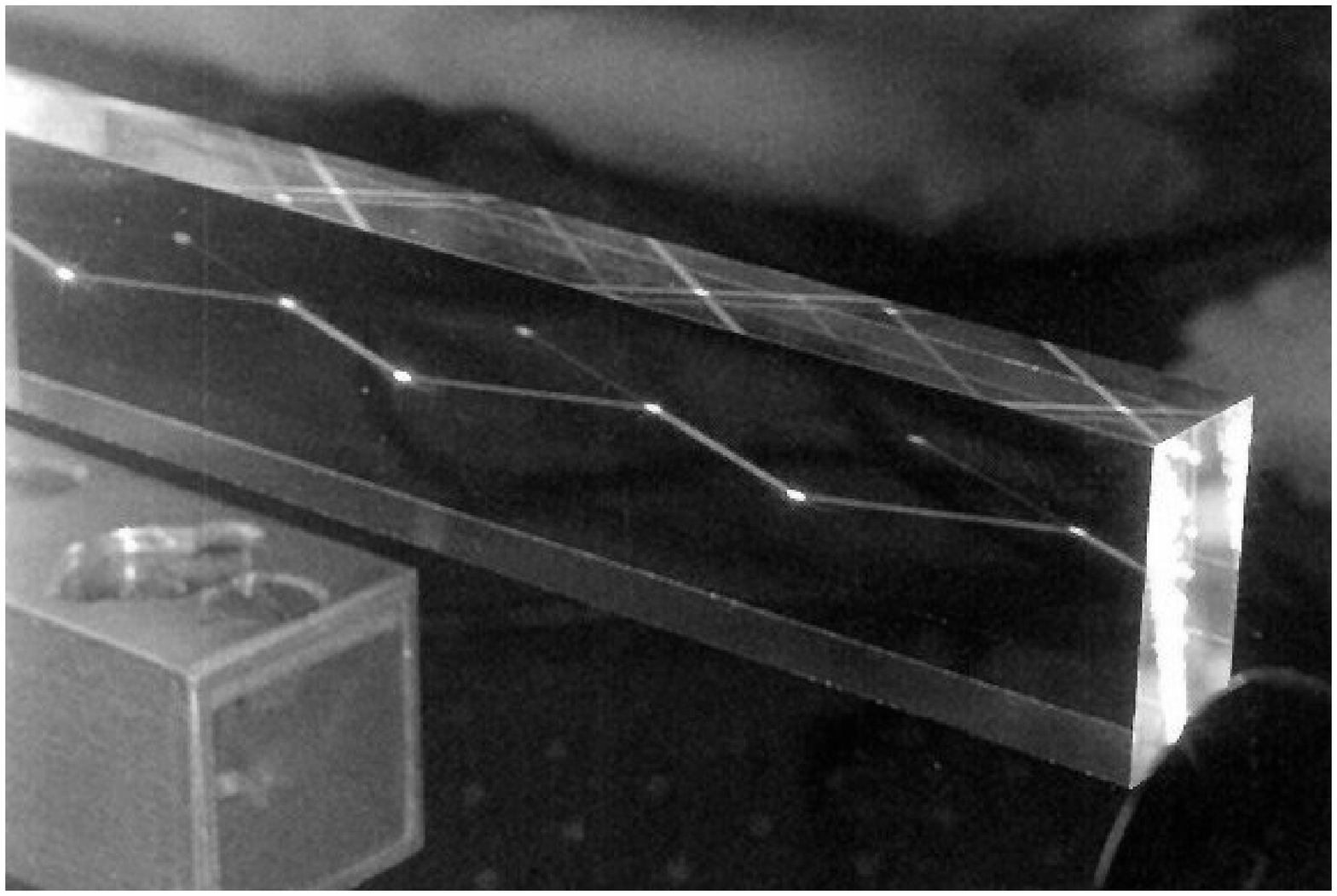}}
\caption{		
\label{Fig:13}
A laser beam bouncing in the fused silica bar during the
relative internal reflection coefficient measurement~\cite{10}.
}\end{figure}

The calorimetric method described above is too time-consuming to be
used for quality assurance measurements  of many bars, or for
measurements of multiple points of a single bar. 
For these types of measurements, we developed an automated setup,
shown in Figure~\ref{Fig:12}~\cite{10}, which makes use of the
vertical polarization of the Liconix 4214NB HeCd laser beam.
The bar can be scanned through the laser beam using an actuator 
controlled by a Macintosh computer running LabView.
The transmission is measured along the axis of the bar.
The reflection coefficient is measured using typically 
53 bounces within the bar (see Figure~\ref{Fig:13}).

A $\lambda$/2 retardation plate converts the vertical polarization
into horizontal polarization so that, when the laser 
enters the bar at the Brewster angle (see Figure~\ref{Fig:12}), 
the intensity of the reflected beam image is close to zero.
This simplifies the overall problem by allowing the measurement of
only two intensities ($\int_0, \int_2$) rather than the five required
in the absolute measurement. 
Another advantage of the automatic method is that it allows a 
scan of about 75 grid-points across a large surface area in less 
than one hour.

The computer-controlled setup measures the relative internal reflection
coefficient, which can be normalized by comparison with a reference
bar whose absolute reflection coefficient has been measured 
manually by the calorimetric method described above.

\begin{figure}
\centerline{\includegraphics[width=11cm]{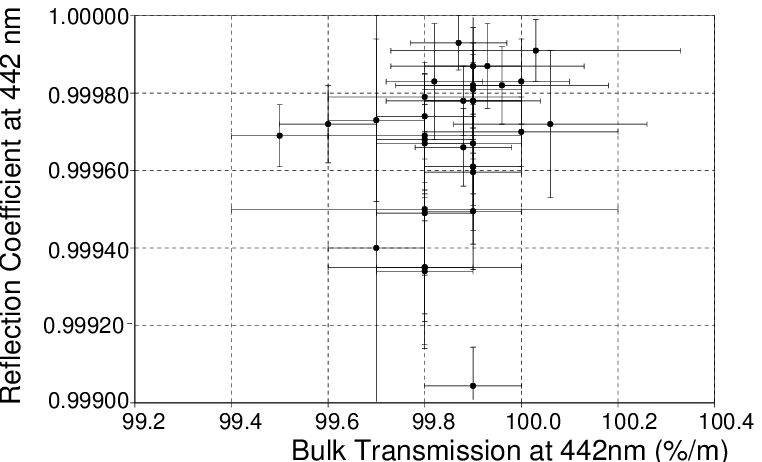}} 
\caption{		
\label{Fig:14}
Measurements of transmission and relative internal
reflection coefficient at a wavelength of 442\ nm~\cite{11}.
}\end{figure}

Prior to the reflection coefficient measurements reported in this
section all bar surfaces were thoroughly cleaned using alumina powder.  
This was done by gently rubbing the bar surfaces with 0.3$\mu$m alumina
powder, which was mixed with clean water in a 1:10 ratio, ensuring that
surface chemical contamination was minimized. 
Cleaning was essential for optimum results. 
The reflection coefficient of a randomly picked bar was measured
before and after being cleaned with the alumina powder method.
The value of the reflection coefficient increased by ~0.0005
after the cleaning.
During DIRC production we found that the bar manufacturer, Boeing
Co.~\cite{boeing}, usually delivered the bars to SLAC with very clean
surfaces. 
Therefore a simpler cleaning method, where the bars were swiped
with acetone and alcohol, was sufficient to restore the optimum
reflection coefficient. 
Throughout the construction phase we monitored the bar
cleanliness using the relative measurements of the reflection
coefficient~\cite{11,12} normalized to selected reference bars.
The average relative internal reflection coefficient of the bars
used in the DIRC construction is 0.9997$\pm$0.0001, the average
transmission is 99.9$\pm$0.1\%/meter at 442\ nm, and  98.9$\pm$0.2\% at
325\ nm.  
Figure~\ref{Fig:14} shows an example of measurement results at
442\ nm~\cite{11}. 

\subsection{Wavelength Dependence of the Reflection Coefficient}
To measure the dependence of \refl\ on the wavelength, we used five
laser wavelengths 
which are provided by four different lasers 
(266\ nm Nanolase solid state YAG laser~\cite{nanolase},
325\ nm and 442\ nm Liconix HeCd 4210N~\cite{liconix},
543\ nm Uniphase 1135 green, 
and 633\ nm Uniphase red~\cite{uniphase}).
The intensity of the lasers is different, influencing the
relative accuracy for each measurement (for example, the Hamamatsu
photodiode currents are ~500-600$\mu$A at 442\ nm, ~20-25mA at 325\ nm,
~10-15$\mu$A at 266\ nm, ~20-25$\mu$A at 543\ nm, and ~500-550$\mu$A at
633\ nm). 
Similarly, laser stability varies, with the 266\ nm Nanolase
being least stable.
			
\begin{figure} 
\centerline{\includegraphics[width=11cm]{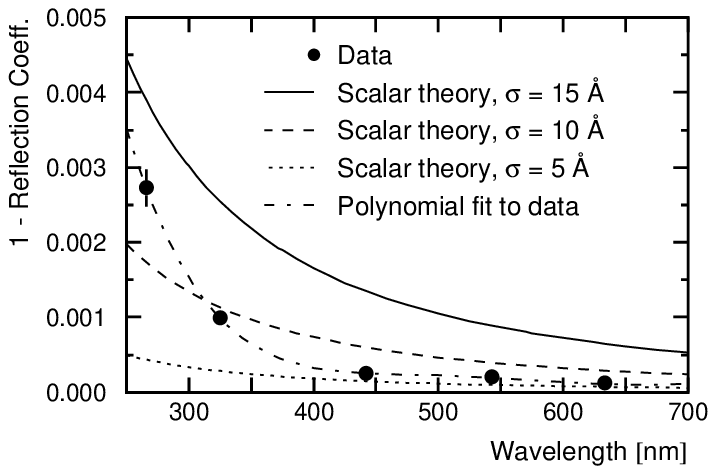}}
\caption{		
\label{Fig:16}
A comparison of the internal reflection coefficient
measurements at five different laser wavelengths and the scalar
scattering theory, assuming only random surface imperfections
with three different surface finishes.
The graph also shows a polynomial fit to data.
}\end{figure}

\begin{table*}
  \begin{sideways}
  \begin{minipage}[b]{.9\textheight}
    \vspace*{3cm}
    \begin{center}
\begin{tabular}{l|l|l|l|l}
Contaminant Candidate & Test Length & 
Temperature & Humidity & Relevance to DIRC \\
\hline
Epotek 301-2 epoxy & 1 day & room & dry & Used for all optical joints\\
Gluing Station & 2 days & room & dry & Exposure to fumes during bar gluing \\
Bar box buttons & 3 days & room & dry & Used in bar box construction \\
Matheson pressure regulator & 10 days & room/ 60$^\circ$ C & dry/wet & 
Used in \babar\ gas system \\
DP-190 Epoxy (translucent) & 31 days & room & dry & 
Used as secondary glue to seal bar boxes \\
DP-190 Epoxy (gray) & 31 days & room & dry & Used during construction of bar boxes\\
Wet Hysol 1C-LV Epoxy & 42 days & room & dry & Used as primary glue to seal 
bar boxes\\
Wet SES-403 RTV & 54 days & room & dry & Not used in DIRC \\
Cured SES-403 RTV & 74 days & room & dry & Not used in DIRC \\
Packing foam for bars & 74 days & room & dry & Used to ship bars 
from Boeing Co.\\
EPDM Gasket & 74 days & room & dry & Used in gaskets for water seal \\
Disogrin O-rings & 75 days & room/ 60$^\circ$ C & dry/wet & 
Used in \babar\ gas system\\
DIRC gas system at \babar\ & 90 days & room & dry & Complete check of
\babar\ gas system\\
Wet Dow Corning 3145 RTV & 116 days & room & dry & Construction of 
DIRC construction clean room \\
Teflon sample & 116 days & room & dry & Not used in DIRC \\
Prototype bar box & 117 days & room / 60$^\circ$ C & dry & 
Initial test of bar box cleanliness \\
\end{tabular}
\end{center}
  \end{minipage}
  \end{sideways}
\caption{\label{tab:pollution} Summary of materials tested as contaminants.}
\end{table*}

Figure~\ref{Fig:16} shows the result,
plotted as (1-Reflection coefficient)
vs. wavelength. 
The plotted errors are based on the $rms$ of ten
repetitive trials for each intensity component $\int_i$. 
There are many sources of systematic error that are difficult to
evaluate.
They depend on many factors, such as surface quality variation, 
surface contamination (especially in the far UV region), 
the method of the photodiode ``peaking'', dust, background light in
the room, the lobe effect due to the refraction index variation, etc. 
To evaluate at least some of these systematic errors, the measurements
are repeated many times, while varying the measurement conditions,
such as laser entrance point into the bar, the laser path within the
bar, the number of bounces N (by a few), avoiding bright spots
observed when hitting either chips or the bar supporting buttons, etc.
After a lot of practice, one can obtain quite repeatable results with
an estimated systematic (repeatability)
error of about 0.0003 to each measurement. 
Figure~\ref{Fig:16} also
shows three curves based on scalar reflection theory~\cite{14},
assuming the fused silica surface finish of 5, 10, and 15 \AA . 
The internal reflection coefficient is consistent with a surface finish
between 5 and 8 \AA\ ($rms$) in the wavelength region between 450 and 
650\ nm. 
The direct measurement by the bar manufacturer Boeing
Co.~\cite{boeing} found that the surface finish was better than 5 \AA\
for all bars, consistent with our measurement of \refl.
However, the measured reflection coefficient falls significantly below
350\ nm, more than predicted by scalar theory. 
This discrepancy is not well understood, although a considerable amount 
of time has been spent checking the experimental results at 266 and
325\ nm. 
A possible explanation for the discrepancy in the UV region could 
be chemical contamination, or near surface humidity. 
Figure~\ref{Fig:16} also shows a polynomial
fit to data.  
This fit is used to evaluate the DIRC efficiency and
mean wavelength response, which is shown in Figure~\ref{Fig:5}~\cite{5}.

\subsection{Bar Surface Contamination Tests}
Extensive contamination tests are performed to see if the internal
reflection coefficient can be degraded by materials used for the
construction of the DIRC bar boxes, gluing, during general handling
procedures, or the DIRC gas system~\cite{15}.
In these tests sample bars are exposed for extended periods of time
to potential contaminants at room and elevated temperatures with and
without added humidity. 
The bar reflection coefficient is either measured continuously
throughout the tests or at regular intervals if the bar is
not accessible during the test.
None of the materials, summarized in Table~\ref{tab:pollution}, causes
any significant decrease of the internal reflection coefficient 
with a measurement precision of about 0.0003.

%% file: radiation.tex
\section{Radiation Hardness of Fused Silica and Optical Glues}
\label{sec:rad}
It is expected that the DIRC will be exposed to 0.5-1\ krad/year during
normal operation in its 10 year lifetime.
Since additional radiation exposure can occur during machine tuning,
it seems reasonable to
require that both the fused silica and the optical glue be
resistant up to at least 10-15\ krad.

\subsection{Study of Radiation Damage of Fused Silica}

\begin{figure}
\centerline{\includegraphics[width=11cm]{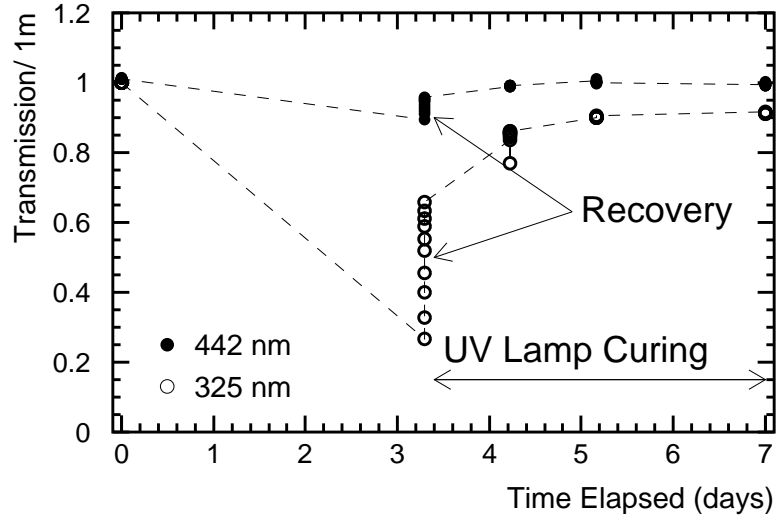}}
\caption{
\label{Fig:17}
Transmission loss at 325\ nm due to radiation damage and
recovery after ``curing'' with UV light for
natural fused silica Vitreosil-F.
Between days 0 and 3, the sample was exposed
to $\sim$7\ krad.
At day 3, rapid curing was observed using a UV laser, but
only within the beam spot of that laser.
From day 3 to 7, slower curing was observed throughout the bulk
of the sample.
}\end{figure}

The initial DIRC prototype tests~\cite{16,17} used
Vitreosil-F~\cite{qpc2}, which is a natural
fused silica material, produced from naturally occurring crystalline
quartz.
Although the optical performance of this material is excellent,
radiation damage tests show that it
lost $\sim$80\% of transmission per meter
at 325\ nm after a radiation dose of only $\sim$7\ krad~\cite{13}
(see Figure~\ref{Fig:17}).
Two other natural fused silica materials
(JGS3-IR~\cite{jgs} and T-08~\cite{heraeus2})
demonstrate similarly rapid deterioration.
With the Vitreosil-F sample, tests were done to determine whether
the radiation damage could be ``cured'' with
photo-bleaching\footnote{Photo bleaching is the use of a high
intensity UV laser or lamp to restore the lost transmission due to the
radiation damage~\cite{18}.}, and/or a heat
treatment.
Figure~\ref{Fig:17} shows that a UV laser 
causes recovery of some of the transmission, but only in
a region close to the beam~\cite{13}.
A strong UV lamp is able to cure the transmission loss over
a larger region.
However, several days are required for this process.
Since such a ``curing procedure'' would be
highly impractical to implement in \babar,
we use the more radiation-resistant, but also
more expensive, synthetic materials.

We conduct two types of radiation damage
studies on the natural and synthetic fused silica samples:
(a) using short samples
(10-20 cm), where the transmission is typically measured with a
monochromator, and
(b) using long samples ($\sim$1\ m), where the
transmission is measured with a HeCd laser (Liconix 4214NB).
In the first type of tests, the transmission
can be measured over a wide range of wavelengths, but with limited
precision.
The second type of test gives more precise measurements,
but at only two discrete wavelengths (325 and 442\ nm).

\label{sec:synthrad}
\subsubsection{Radiation Source}
We use the Co$^{60}$ source facility available at SLAC to irradiate
the bars~\cite{13,19}.
Its activity is about 14.1 Curie.
A short sample (30 cm) can be placed very close to the source and
be irradiated with a dose up to $\sim$50-60\ krad per day.
The Co$^{60}$ source produces two major
photon emission lines, 1.173 and 1.333 MeV.
There are some Compton
electrons coming from the well concrete walls; 
they contribute about 10\% to the total dose.
The radiation dose was estimated using TLD dosimeters.
The measurement is calibrated using
(a) a theoretical estimate based on the geometry and known source
activity corrected for its lifetime, (b) an ion chamber measurement,
and (c) an opti-chromic dosimeter measurement.
The measurements agree to within $\sim$10 \%.

\subsubsection{Monochromator Transmission Studies of Natural Fused
  Silica}
\begin{figure}
\vspace*{-12mm}
\centerline{\includegraphics[width=11cm]{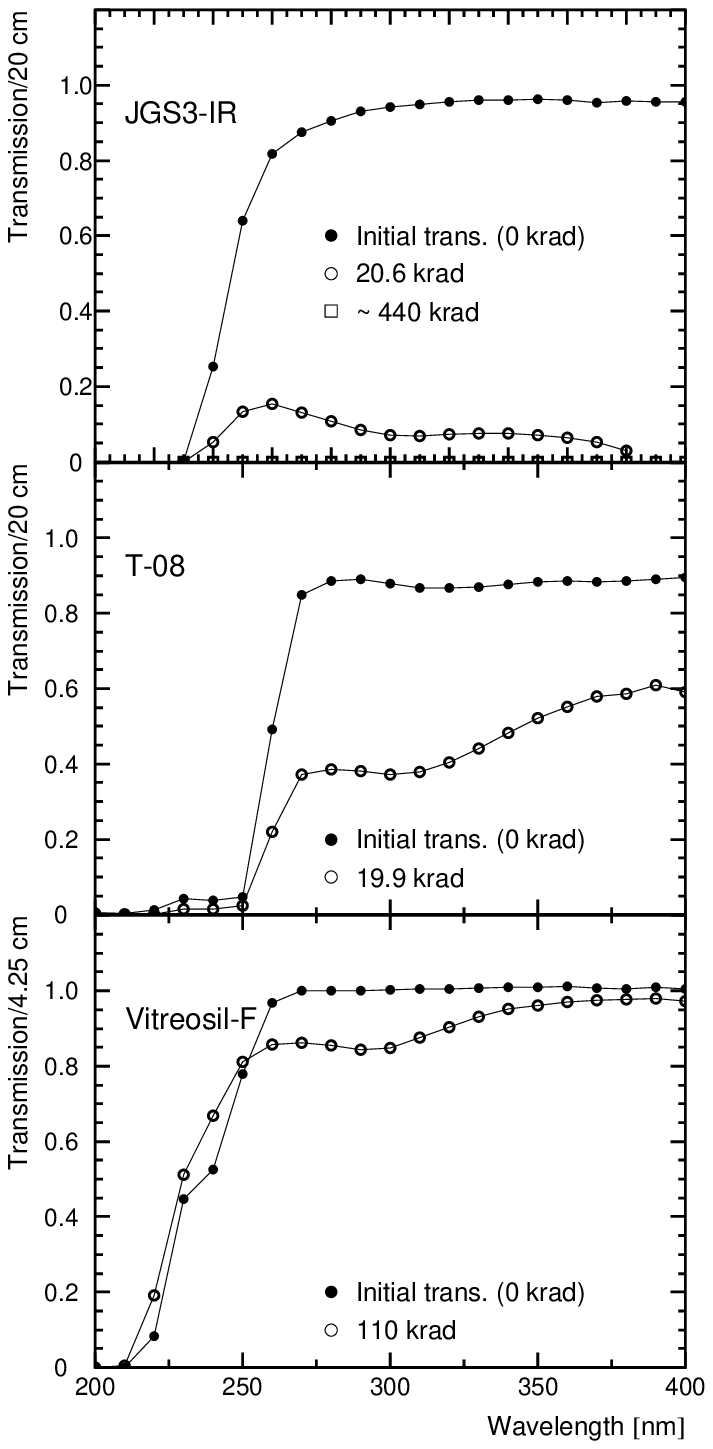}}
\caption{
\label{Fig:18}
Transmission curves for several types of natural fused silica
before and after irradiation~\cite{13}.
}\end{figure}

The transmission of the fused silica samples is measured as a 
function of wavelength in a monochromator, before and after exposure
to the radiation source.
A correction is made for the Fresnel reflections at the entrance
and exit surfaces.
Since the polish of these surfaces, performed at SLAC, is of moderate
quality, some additional light (typically 2\%)
is lost due to scattering.
However, the scattered light should be roughly the same
before and after irradiation, so this does not effect the
accuracy of the radiation damage measurements.
Figure~\ref{Fig:18} shows the results of the monochromator studies of
radiation damage for the three types of natural fused silica that
were studied.
These measurements have systematic errors at a level of 2-3\%. 
For all three types, the transmission is seriously degraded
after a dose of only 20\ krad.
The DIRC with the 5-meter long
bars would not function at all given such transmission losses.

\begin{table*}
  \begin{sideways}
  \begin{minipage}[b]{.9\textheight}
    \vspace*{3cm}
\begin{center}
    \hspace*{1cm}
\begin{tabular}{l|l|l|l|l|l}
Material & Manufacturer & Type & Visual Change & Radioluminescence & Transmission Loss \\
\hline
Vitreosil-F & TSL & Natural &  No (100\ krad) &
No (100\ krad) & Severe (7\ krad) \\
\hline
T-08 & Heraeus Amersil & Natural & Yellow (330\ krad) &
Yes (330\ krad) & Severe (10\ krad) \\
\hline
JGS3-IR & Beijing Institute &
Natural & Brown (400\ krad) & No (400\ krad))&Severe (20\ krad) \\
\hline
Suprasil & Heraeus Amersil & Synthetic & No (280\ krad) & Yes (280\ krad) &
Small (280\ krad) \\
\hline
JGS1-UV & Beijing Institute & Synthetic & No (650\ krad) & No (650\ krad) &
No (650\ krad) \\
\hline
Spectrosil 2000 & TSL & Synthetic & No (180\ krad) & No (180\ krad) &
Small (180\ krad) \\
\hline
Spectrosil B & TSL & Synthetic & No (254\ krad) & Yes (254\ krad) & Small (254\ krad) \\
\end{tabular}
\end{center}
  \end{minipage}
  \end{sideways}
\caption{\label{tab:raddam}Effects of radiation damage for various types of
natural and synthetic fused silica samples~\cite{13,19}. }
\end{table*}

In addition to transmission loss, visual changes and radio-luminescence
were also observed as consequences of radiation damage.
Table~\ref{tab:raddam} summarizes these observations.
The radio-luminescence observed in the T-08 sample caused monochromator
singles rates to increase by a factor of $\sim$37 after
a dose of \mbox{$\sim$10\ krad}, and by a factor
of \mbox{$\sim$8500} after a dose of $\sim$334\ krad.
Such radio-luminescence is a known property of quartz and fused
silica~\cite{20}.
Neither Vitreosil-F nor JGS3-IR samples
showed any noticeable radio-luminescence after the irradiation.
However, these measurements are not sensitive to radio-luminescence with
a very short time constant, since the first measurements are usually
done at least 30-60 minutes after the sample is removed from the
radiation well.

\subsubsection{Monochromator Transmission Studies of Synthetic Fused Silica}
\begin{figure}
\centerline{\includegraphics[width=8cm]{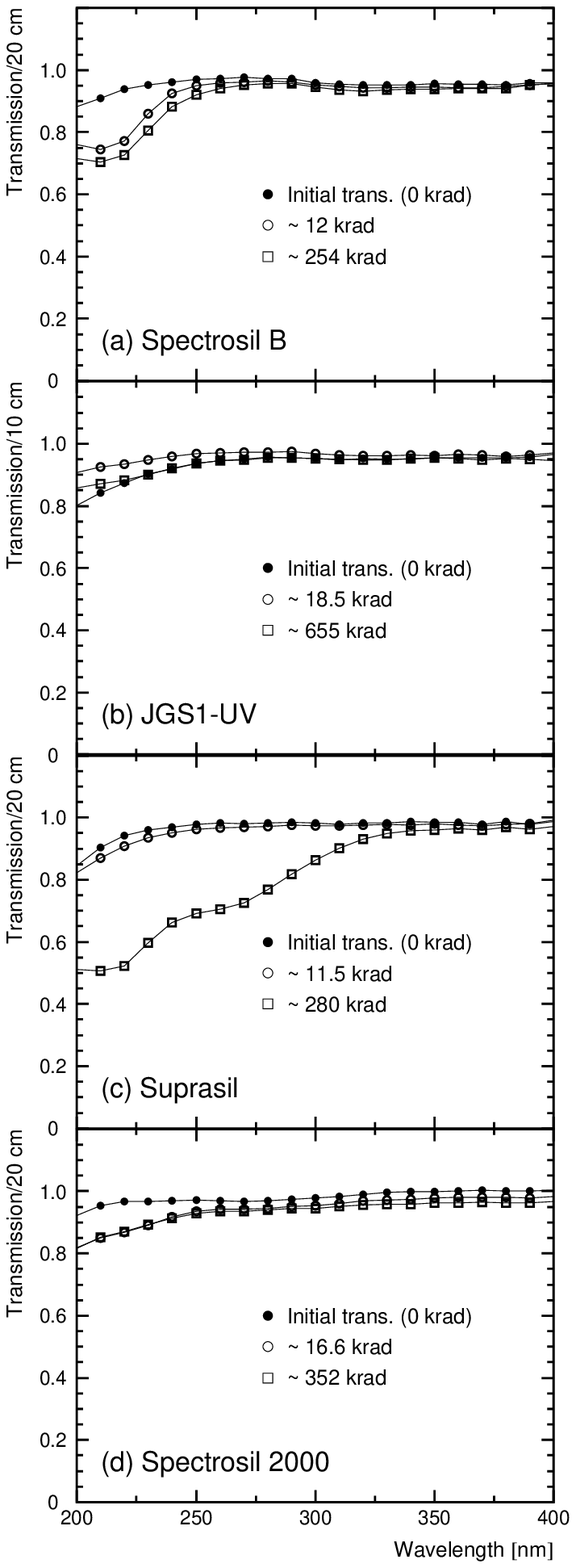}}
\caption{
\label{Fig:20}
Transmission curves for several types of synthetic fused
silica before and after irradiation~\cite{19}.
}\end{figure}

It is clear that natural fused silica is not a good candidate for the
\babar\ experiment, and therefore, we have studied the properties of
synthetic fused silica material.
We have tested four short synthetic fused silica samples.
Spectrosil B~\cite{qpc} (2 cm dia. x 20 cm long sample),
JGS1-UV~\cite{jgs2} (2 cm x 2 cm x 10 cm sample),
Suprasil Standard~\cite{heraeus} (2 cm dia. x 20 cm long sample),
Spectrosil 2000~\cite{qpc}
(2 cm x 2 cm x 20.7 cm long sample).
Figure~\ref{Fig:20} shows the results of these studies.

\begin{figure}
\centerline{\includegraphics[width=11cm]{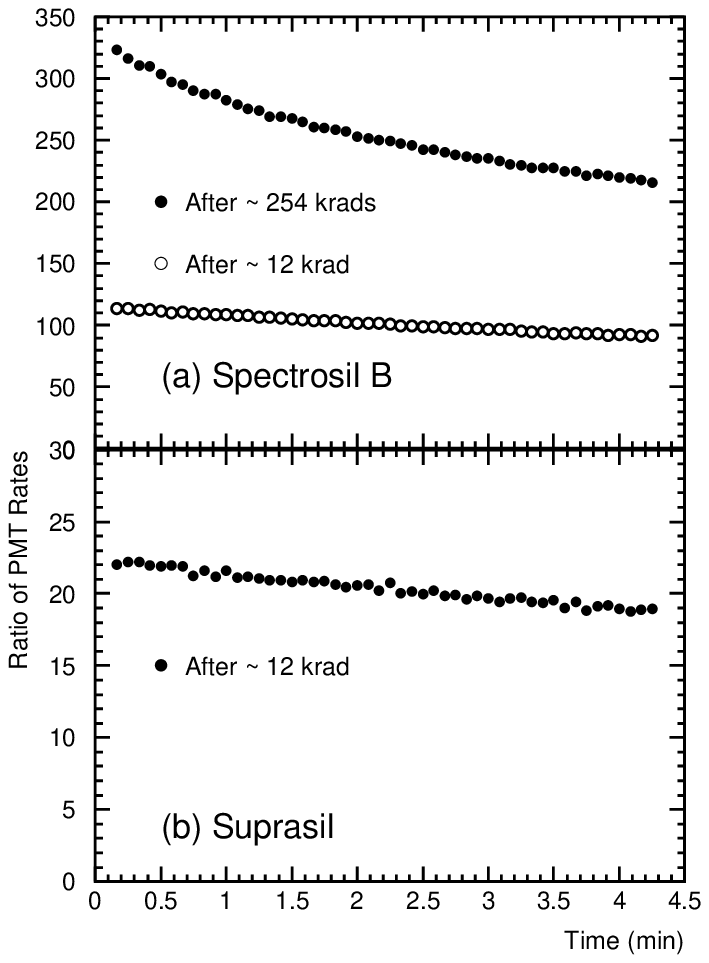}}
\caption{
\label{Fig:21}
Radio-luminescence of synthetic fused silica, as seen
as the ratio of PMT rates in the monochromator~\cite{19} for
(a) Spectrosil B and
(b) Suprasil.
The measurements are started half an hour after removing the samples from
the Co$^{60}$ well.
}\end{figure}

Figure~\ref{Fig:20}(a) shows that Spectrosil B is
radiation hard above 280\ nm.
In the far UV region below 240\ nm, we
observe some radiation sensitivity.
However it is still within
acceptable levels for the \babar\ experiment.
This sample showed 
strong radio-luminescence after radiation exposure, as one can see in
Figure~\ref{Fig:21}(a).
The photonic activity was estimated with the monochromator's
PMT, and was about 100 times more than normal after $\sim$12\ krad, and was
decaying at a rate of about 10\% after 2-3 minutes.
The time interval
between the end of the irradiation and the beginning of the
transmission measurement was typically 30-40 minutes.

JGS1-UV  synthetic material (Figure~\ref{Fig:20}(b)) is very radiation
resistant.
In fact, it is among the best samples we have tested. This
sample did not show any radio-luminescence after radiation exposure.

Suprasil Standard synthetic (Figure~\ref{Fig:20}(c))
material shows the largest radiation damage for all synthetic materials
we tested.
The material is radiation hard for small
doses up to 10-20\ krad.
However, when the dose is increased to $\sim$280
krad, we see a substantial transmission loss below
$\sim$340\ nm.
Furthermore, Figure~\ref{Fig:21}(b) shows that this particular
material produced radio-luminescence after radiation exposure.

Figure~\ref{Fig:20}(d) shows the transmission of the Spectrosil
2000 sample for several irradiation doses.
This particular
synthetic material shows good resistance to radiation.
The material lost 2-3 \%/meter of transmission above 260\ nm after
a dose of $\sim$17\ krad.
In the far UV region below 240\ nm, we observe larger losses.
However, it is well within the requirements of the
\babar\ experiment.
This sample did not show any radio-luminescence after radiation
exposure.
Table~\ref{tab:raddam} summarizes all radiation damage
tests for all of the types of fused silica that we tested.

\subsubsection{Study of Radiation Damage of Long Synthetic Fused Silica
Samples}
\label{sec:synthradbar}
In order to verify the results obtained in the monochromator
with short samples (20 cm) of synthetic fused silica, similar
measurements were done with long ($\ge$ 100 cm) samples.
These measurements use the laser scanning system described in 
section~\ref{sec:transrefl}~\cite{12}.
This setup is capable of measuring the
average transmission with a systematic error of about 0.2\%.

We have tested two synthetic fused silica materials: (a) a Suprasil
Standard rod, 100 cm long and 26 mm in diameter, and (b) a
Spectrosil 2000 rod, 137 cm long and 55 mm diameter~\cite{19}.

The long fused silica samples are irradiated using a Co$^{60}$ source 
($\sim$1400 Curie) at LBNL. 
The bar is placed
perpendicular to the geometric axis of the source.
At a distance of 1 meter from the
source, one can achieve a dose of $\sim$38\ kR per day.
The radiation dose was measured with a Victoreen dosimeter measuring
the exposure in Roentgen units (R)\footnote{In Roentgen units (R),
1 R = 2.58 $10^{-4}$ Coul kg-1.
The TLD dosimeter measures the absorbed dose in Rad units.
The conversion between rad and R depends on the material.
In tissue, it is 1 rad = 1.139 R .}.
Therefore, in the following, the radiation doses will be given in
terms of exposure dose, using Roentgen units.

\begin{figure}
\vspace*{-2mm}
\centerline{\includegraphics[width=11cm]{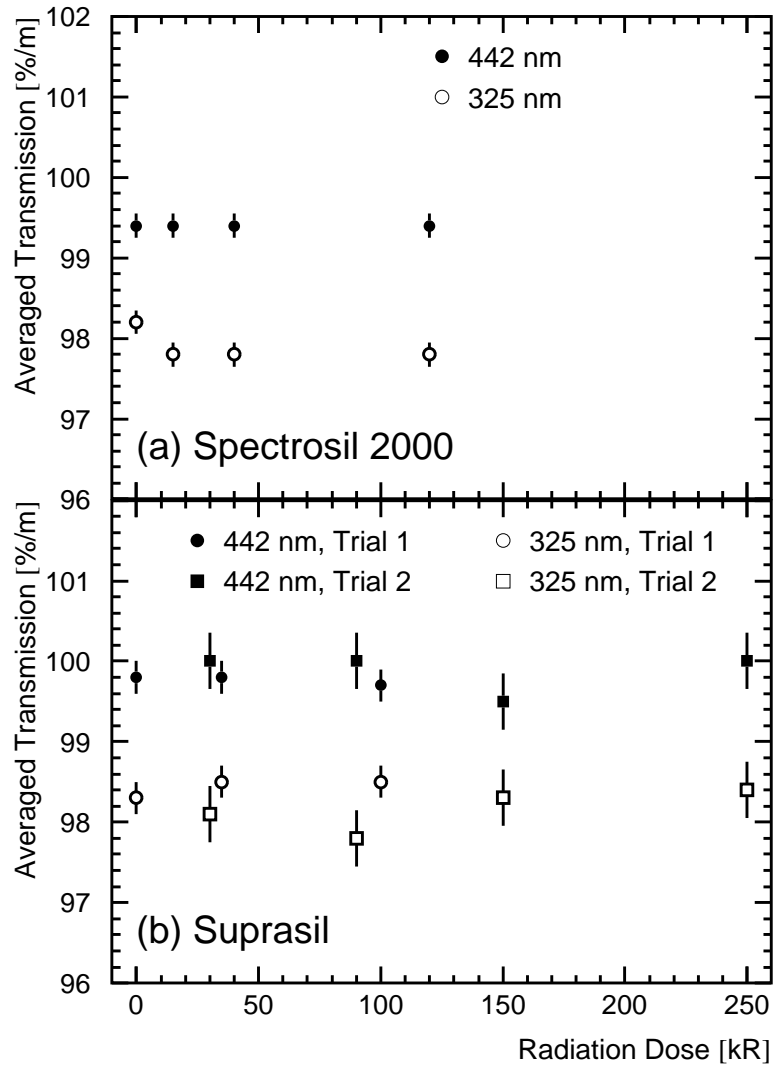}}
\caption{
\label{Fig:25}
Radiation damage of the synthetic fused silica.
(a) shows the results for Spectrosil 2000 (137 cm long) sample.
(b) shows the results for Suprasil~\cite{19}.
}\end{figure}

Figure~\ref{Fig:25}(a) shows the transmission measurements for
Spectrosil 2000.
The material loses $\sim$0.5\%/m of transmission at 325\ nm after 
15\ kR.
Increasing the dose further does not result in additional
transmission loss.
No effect is observed at 442\ nm for all radiation
doses.
Figure~\ref{Fig:25}(b) shows the transmission measurements for Suprasil
Standard.
No significant transmission loss is observed at 325\ nm or
442\ nm.
This result is consistent with the radiation damage observed in
the small samples, but does not extend into the UV.

Based on the radiation test results, we have found that the QPC
Spectrosil 2000 fused silica material satisfies requirements for the
radiation hardness needed to operate successfully for 10 years at
\babar.

It is interesting to note that the same samples of fused silica
described in this section were used in subsequent test performed
during R\&D for the E158 experiment at SLAC.
The Spectrosil 2000 material used in DIRC was found to be radiation
hard to photon doses of several Mrad and neutron doses of
more than 30\ krad~\cite{e158}.

\subsection{Study of Radiation Damage in Optical Glues}
It is also necessary to test the radiation hardness of the optical
glues, which could fail either due to radiation or background induced
light in the bar.
We have tested five glue candidates: (a)
Epotek 301-2 optical epoxy, which is the glue actually used for gluing
of all DIRC fused silica bars, (b) SES-403 RTV, (c) SES-406 RTV, (d)
KE-108 RTV~\cite{shin-etsu},
and (e) Rhodorsil-141 RTV~\cite{rhone}.
All optical glue samples
are tested using the SLAC Co$^{60}$ source facility.

The following procedure is used:
a glue sample is placed between two small fused
silica plates, each 4mm thick.
This particular fused silica is radiation resistant up to at
least 70\ krad. 
It cuts the transparency at around
170\ nm, so that it does not influence the glue studies in this note.
The glue samples are placed in the monochromator to check the initial
transmission, then irradiated with a dose of about 60-70\ krad, and
then rechecked for transmission.

\begin{figure}
\centerline{\includegraphics[width=11cm]{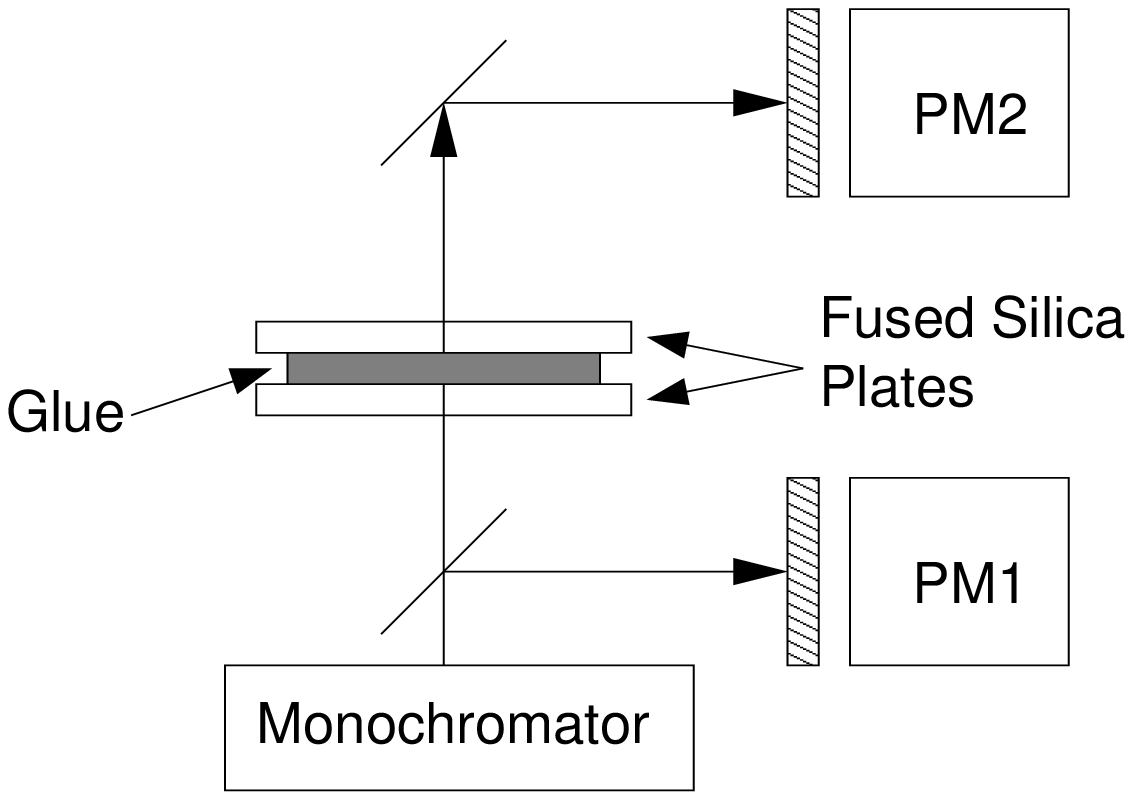}}
\caption{
\label{Fig:gluesetup}
Experimental setup used for the glue transmission tests.
The glue was placed between two fused silica
plates, which were radiation hard.
The glue sample was typically $\sim$25$\mu$m thick~\cite{19}.
Light was measured with two photomultiplier tubes.
}\end{figure}

\begin{figure}
\centerline{\includegraphics[width=10cm]{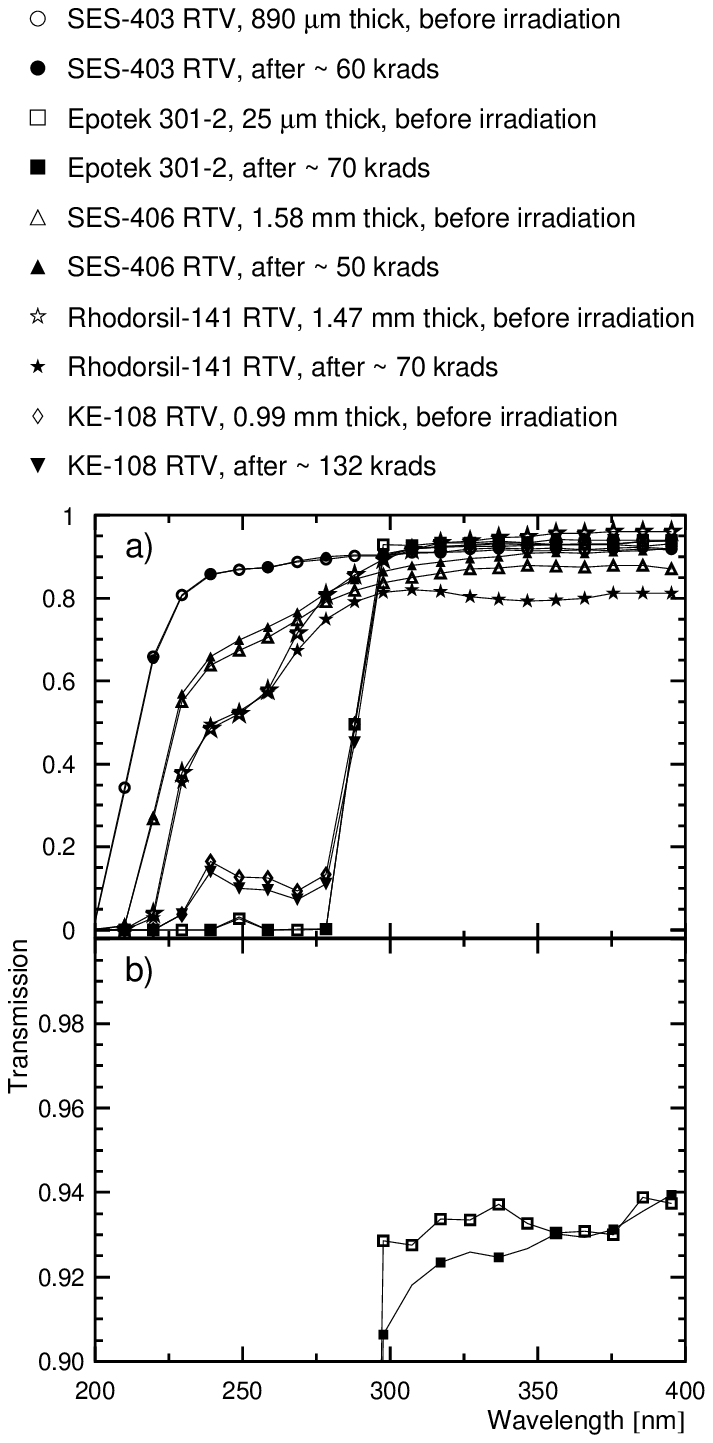}}
\caption{
\label{Fig:27}
Transmission measurements versus wavelength for several
optical glue candidates before and after radiation exposure.
(a) summary of all glues tested.
(b) detailed view of Epotek 301-2~\cite{19}.
The curves have not been corrected for Fresnel reflection.
}\end{figure}

Figure~\ref{Fig:gluesetup} shows the experimental setup used
for the glue transmission studies.
Figure~\ref{Fig:27}(a) shows all the results for all the glues
that were tested.
Figure~\ref{Fig:27}(b) shows the detailed results for
Epotek 301-2 epoxy, which is an optical epoxy with a UV cut-off edge
at $\sim$280\ nm.
One can see a slight transmission loss of about 1\%,
between 300 and 350\ nm, after a dose of $\sim$70\ krad.
The thickness of the Epotek sample,$\sim$25$\mu$m,
is the same thickness typically used for gluing bars together.

Based on these tests, Epotek 301-2 epoxy satisfies requirements 
to operate successfully for 10 years at \babar , and was 
chosen to glue the DIRC bars together.

In addition, glue samples were exposed to a strong photon flux.
No discoloration nor any loss of transmission is observed.

%% file: shims.tex
\section{Reflection from the Fused Silica Bar Supporting Shims and
Mirrors} 
\label{sec:shims}

\begin{figure} 
\vspace*{-2mm}
\centerline{\includegraphics[width=11cm]{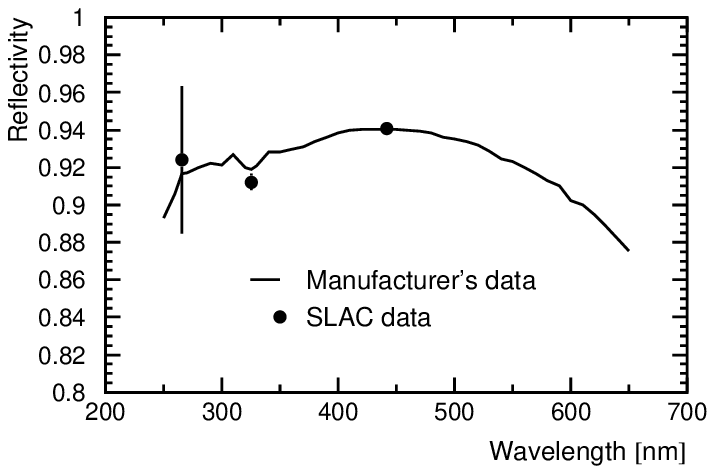}}
\caption{		
\label{Fig:29}
DIRC mirror reflectivity as a function of wavelength is
shown. 
The data are measured with a laser, the mirror is in
air, the laser beam is perpendicular to the mirror, and it is
polarized vertically. 
The mirror manufacturer's data are
measured with a monochromator~\cite{customO}.
}\end{figure}

The mirror~\cite{customO} at the end of each bar is an important part of the DIRC 
enabling the capture of the Cherenkov light going forward into \babar\
(the DIRC PMT readout is on the backward end of the detector).
As a check of the manufacturer's data, we have measured the
reflectivity of these mirrors using a laser and photodiode setup 
similar to that described in section~\ref{sec:transrefl}.
Figure~\ref{Fig:29}
shows the DIRC mirror reflectivity as a function of wavelength. 
The manufacturer's data are consistent with our data, which are taken
at laser wavelengths of 266, 325, and 442\ nm.

Another important requirement of the DIRC is that the bars must
held fixed within the bar box.
This is done with a number of ``buttons'' that support the bars and
``shims'' that keep the bars separated from each other.
The materials used for these supports should ideally have high
reflectivity.
Furthermore, since there can be relative motion between the bars and the 
bar box, the button material should not scratch the surface of the 
fused silica.

\begin{figure} 
\vspace*{-12mm}
\centerline{\includegraphics[width=11cm]{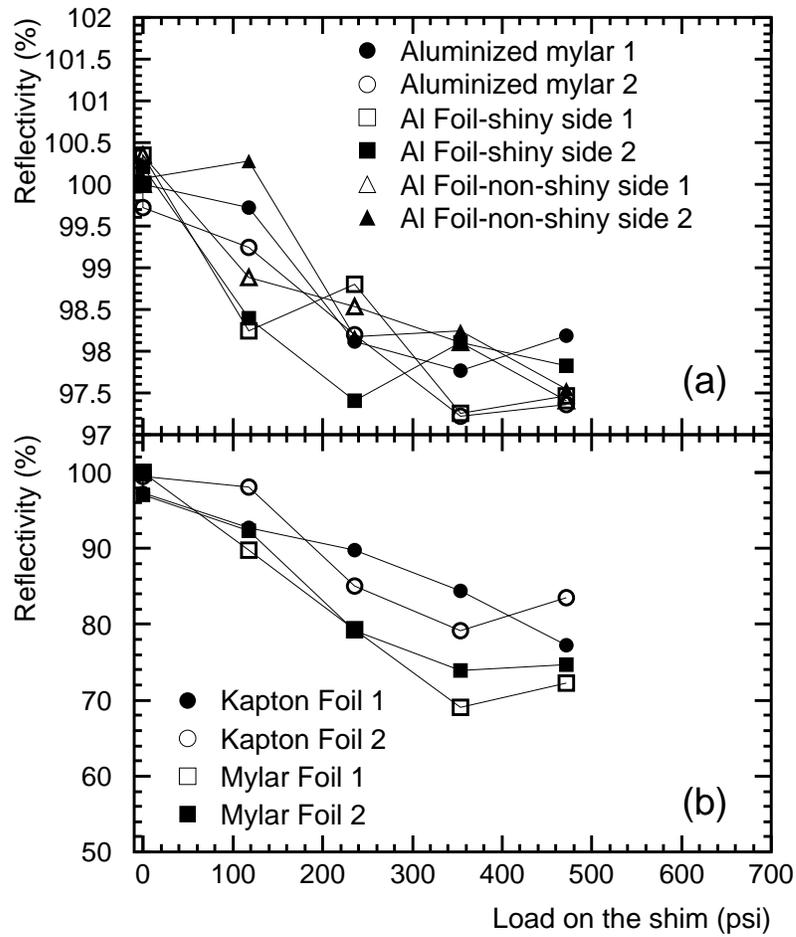}}
\caption{		
\label{Fig:31}
Single bounce reflectivity for several different types of 
shim materials, which contact the bar at the point of reflection.  
We used non-shiny aluminum shims to separate the bars in DIRC. 
The measurements were performed with
blue laser light (442\ nm) at an incident angle of 
21.5$^\circ$~\cite{5}.
}\end{figure}

The reflectivity of a number of different potential shim materials is 
measured using a setup similar to that described in
section~\ref{sec:transrefl}. 
Table~\ref{tab:fig30} shows the results of these measurements and 
indicates that, for example, Nylon creates substantial
reflection losses. 
This is true for all of the soft plastic
materials tested such as Teflon or Rubber. 
The harder plastic
materials, such as Kapton or Mylar, perform better. 
Metals, such as Aluminum, perform even better. 
Table~\ref{tab:fig30} indicates that the
reflectivity of any material depends on the pressure with which the
material is pressed against the fused silica bar. 
To quantify this
better, we constructed a shim press, which is calibrated to express
the force per area in psi units. 
Figure~\ref{Fig:31}(a) 
shows the reflection losses
with Aluminum shim foils as a function of the load. 
One can see that
the loss is less than 1\% for typical practical loads. 
For comparison,
Figure~\ref{Fig:31}(b) shows that the uncoated Kapton 
or Mylar shims would be a poor choice.

\begin{table}
\begin{center}
\begin{tabular}{l|cc}
Material & \multicolumn{2}{c}{\hspace*{-6mm}Reflectivity} \\
\hline
& \multicolumn{2}{c}{Contact Pressure\hspace*{6mm}}\\[1mm]
& Light & Firm \\
\hline
Kapton & 1.000 & 0.994 \\
Mylar & 1.000 & 0.985 \\
Gloves (rubber) & 0.985 & 0.938 \\
Nylon button & 0.989 & 0.949 \\
Al (shiny) & 1.000 & 0.990 \\
Al (non-shiny) & 1.000 & 0.990 \\
Fused Silica (polished) & 1.000 & 0.996 \\
\end{tabular}
\end{center}
\caption{\label{tab:fig30} The single bounce reflectivity from 
various materials,
which are in contact with the bar at the point of reflection. 
These qualitative results motivated the quantitative study
shown in Figure~\ref{Fig:31}~\cite{5}.}
\end{table}

There is an interesting physical explanation of these tests: a portion
of the internally reflecting wave extends beyond the fused silica
boundary and probes the medium on other side. 
As a result, the wave may not be totally internally reflected.
This effect gets larger as the
laser wavelength decreases and is also sensitive to the proximity to
the critical angle. 
This effect is used for commercial pollution
studies, usually in the infrared wavelength region. 
In our case, as
the force on the shim gets larger, this effect increases, resulting in
the reflection coefficient decrease.  
Based on these tests, we chose Aluminum as the material 
for shims between the bars.

To select the material for the supporting buttons separating the bars
and aluminum bar box we performed tests in which buttons were rubbed
several thousand times across the bar surface.
Since the area in which the support buttons are in contact with the
bar surface is very small, the single bounce reflectivity of 
the material is less important than avoiding scratching the bars.
Therefore, Nylon is chosen as button material because it does not
damage the surface of the bar. 
		

%% file: mech.tex
\section{Fused Silica Bar Mechanical Properties}
\label{sec:mech}

The mechanical bar properties, such as the orthogonality between bar
surfaces and edge damage or chips can clearly influence the quality of
the Cherenkov angle image or the number of photons detected. 
The primary goals of our
setup are to validate the manufacturer's~\cite{boeing}
bar measurement procedures, and to 
provide independent measurements of a subset of bars to 
verify stability in the quality assurance (QA) process.
To perform these QA checks at SLAC, we use a digital
microscope~\cite{8} and several image-treating analysis programs 
based on NIH Image software~\cite{nih}.

\begin{figure} 
\centerline{\includegraphics[width=11cm]{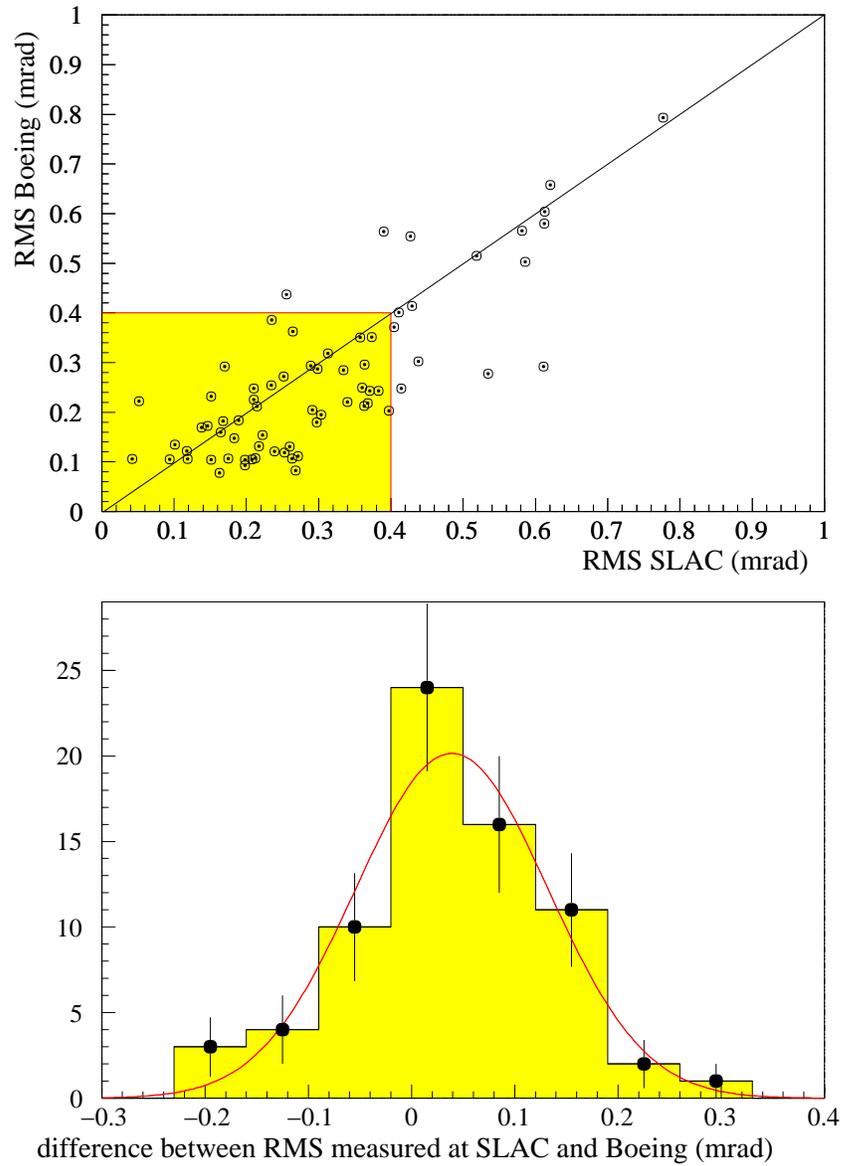}}
\caption{		
\label{Fig:33}
(a) Scatter plot shows the $rms$ measurements made by Boeing
and by SLAC. The rejected bars are outside of the shaded
rectangle. (b) Difference between two measurements of the $rms$ quantity
defined in the text.
}\end{figure}

The first method is based on manual operation, where an operator
follows an edge of the bar visually on the microscope screen, 
clicking on several points along the edge to digitize its
coordinates. 
Such points are then used in a fit to find a straight
line. 
Using two straight lines one could then define the orthogonality
of two surfaces of the bar.

\begin{figure} 
\centerline{
\includegraphics[width=3.5cm]{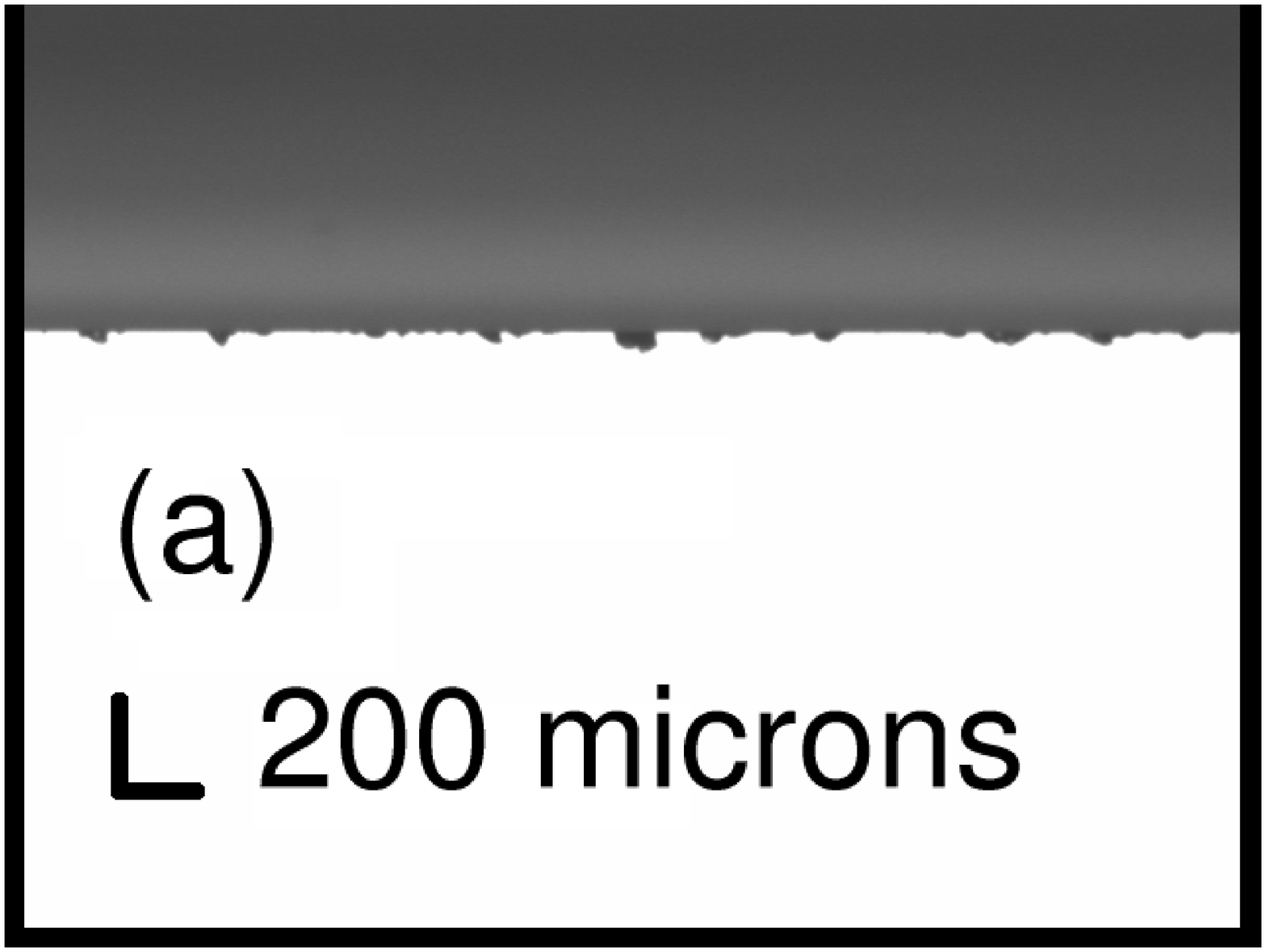}
\includegraphics[width=3.5cm]{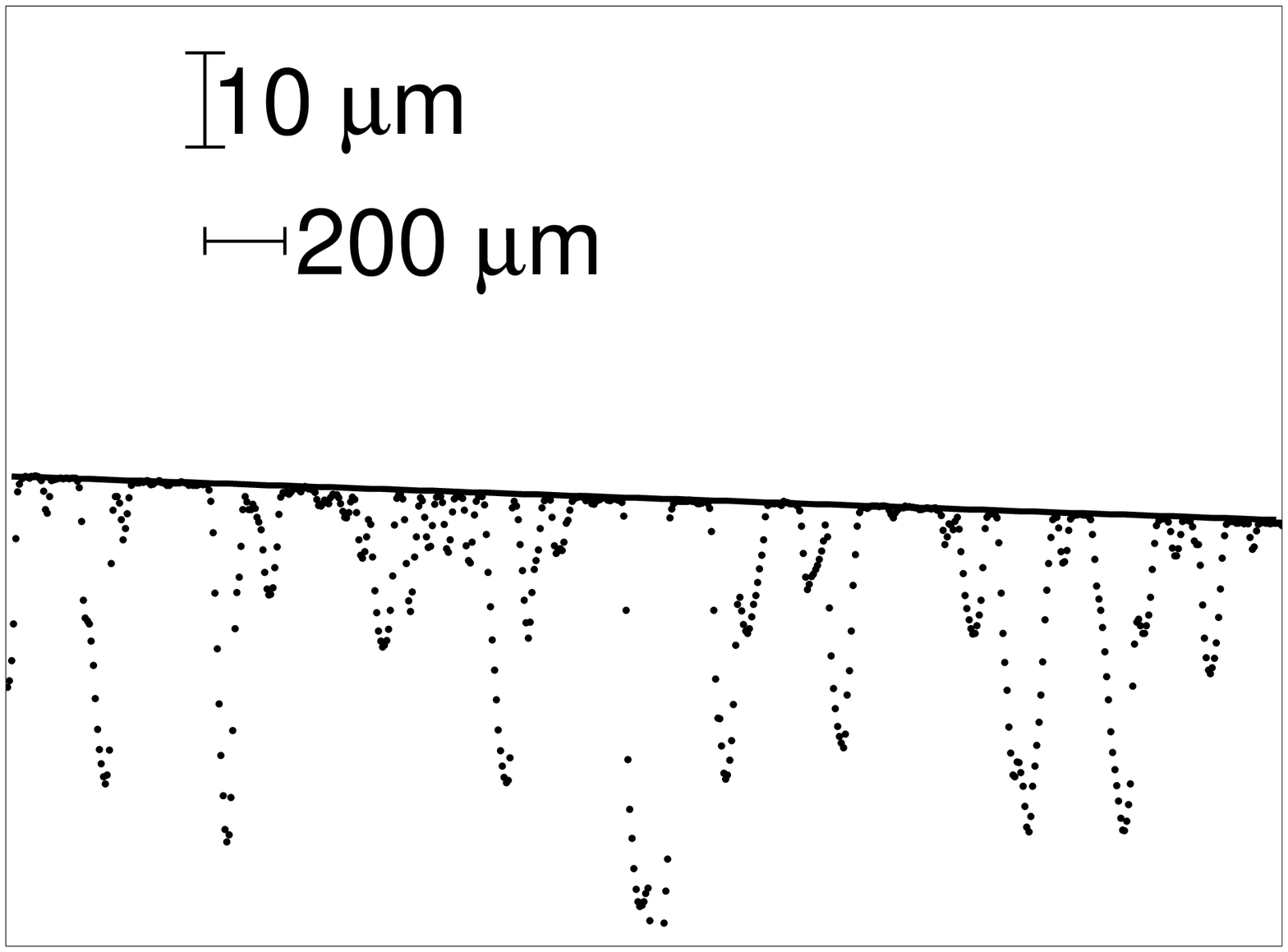}
}
\vskip .1in
\centerline{
\includegraphics[width=3.5cm]{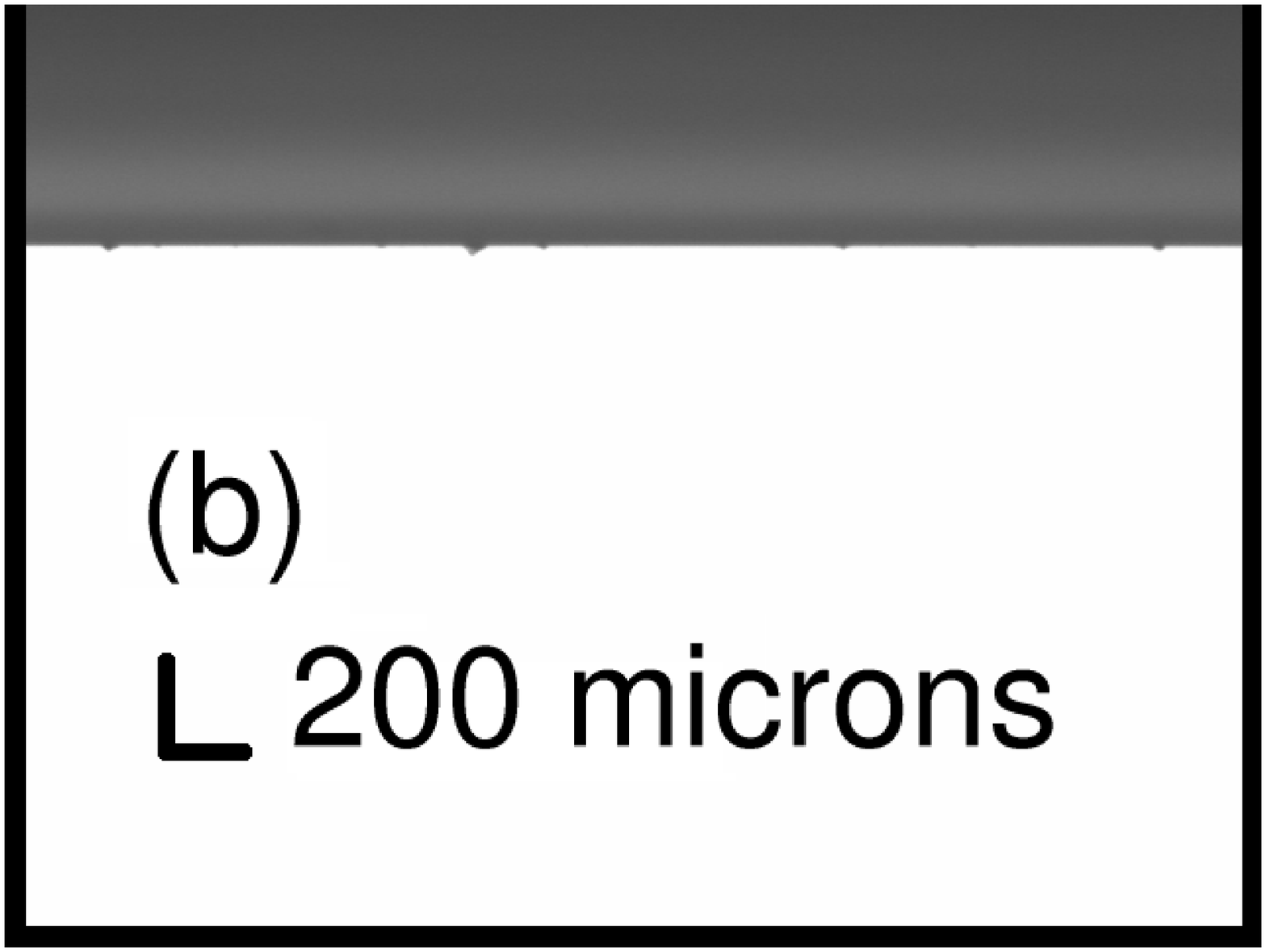}
\includegraphics[width=3.5cm]{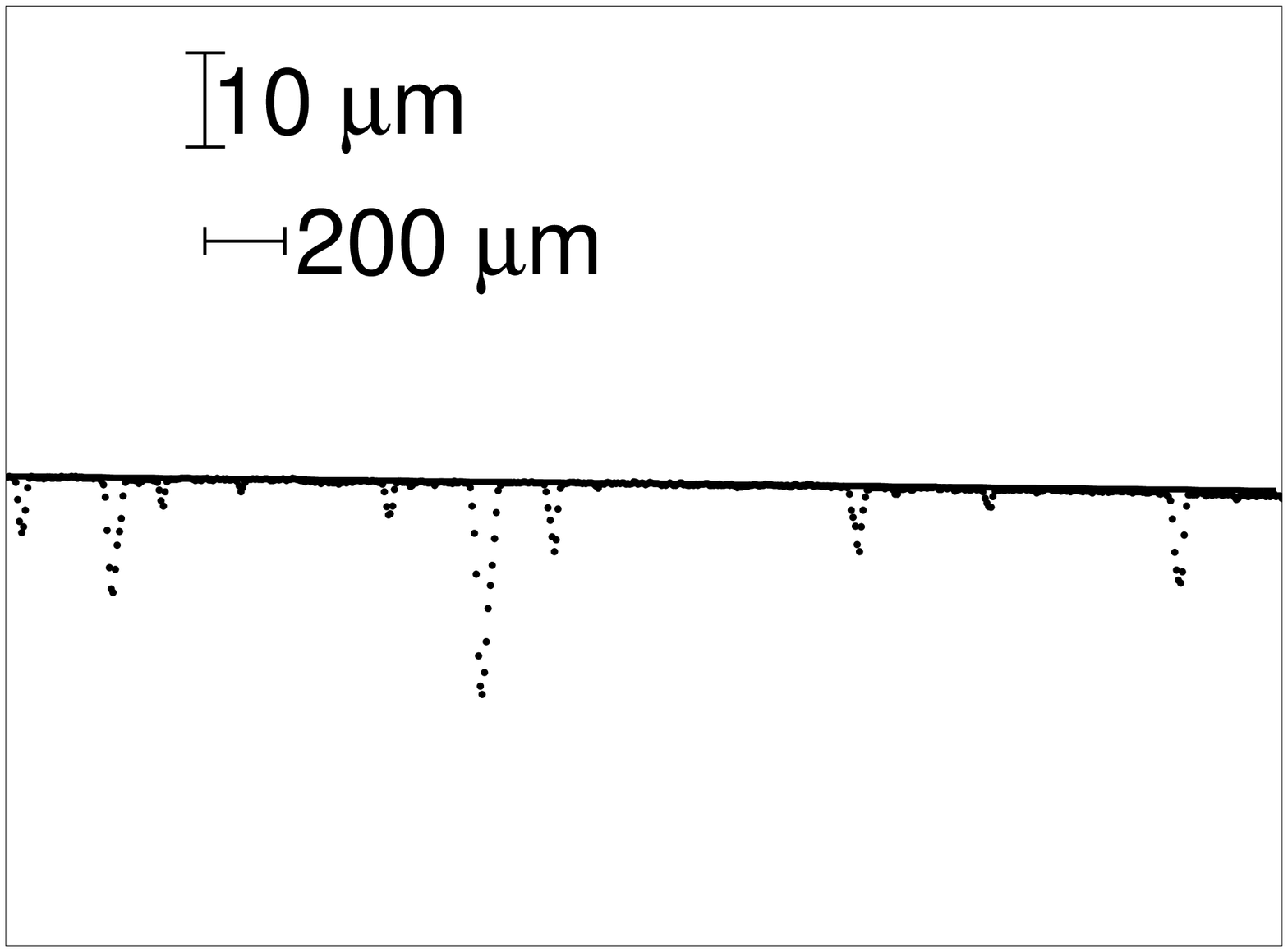}
}
\caption{		
\label{Fig:34}
Examples of (a) a bad bar edge, as it appears on the computer
screen used for the microscope readout (left side), and the digitized
version of the same edge by the off-line software analysis (right
side), and (b) the same for a good bar edge, 
which was typical~\cite{8}. 
Note that even for badly damaged edges, the algorithm finds the
``true'' edge with good accuracy. 
The digitized image has been greatly magnified in the y-direction.
}\end{figure}

The second, algorithmic, approach uses software to analyze a
digital picture of the bar taken with the microscope.
Each picture consists of an array of 640 by 480 pixels, each of
which has 8 bits of gray-scale information. 
The microscope magnification is such that each pixel represents
an area of about 5 microns by 5 microns.
For each picture, the bar edges are found by stepping through each 
row and column of the pixel
matrix and looking for the transition between black and white with a
simple threshold algorithm looking at the gray scale information. 
This method is used to measure the
fused silica bar angles by making an image that includes two edges. 
A typical bar edge consists of several hundred points that are
used to define the edges. 
A special line-fitting algorithm is needed
to reject those points lying on ``chips''. 
We achieve resolution on the edge position of about 1 micron. 

The manufacturer uses a micrometer-based fixture to measure bar angles,
and this method was occasionally calibrated with an auto-collimator
(this method yields a precision of approximately 50\ $\mu$rad). 
Comparing differences in side-to-end and face-to-end angle measurements using
our method and the Boeing method yielded a fitted error of less than
0.15\ mrad, which gave us confidence that the two methods are
consistent and can be used to reject bars that are out of
tolerance. 
The face-to-side angle measurement comparison gives adequate, but 
somewhat worse agreement of about 0.23\ mrad. 
We accept bars to be used
in the DIRC using the following statistical procedure: 
using all four face-to-side angles on both sides of the bar, we
calculate the $rms$ of the eight values. 
The production tolerance is 0.4\ mrad for the $rms$ of
eight values, with a goal of 0.25\ mrad in each angle. 
Figure~\ref{Fig:33} shows an example of the good agreement between
$rms$ measurements  by Boeing Co. and SLAC, which gave us confidence
in the manufacturer's data. 
				
Finally, the digital treatment of the entire image allows a 
quantitative analysis to be performed on the quality of the edges.
Figure~\ref{Fig:34} shows an
example of a good and bad bar edge as seen by the microscope and the
corresponding software. 
Once bar production began, all of the bars easily passed the
requirement that the total area of edge chips be less than 6\ mm$^2$.

%% file: concl.tex
\section{Conclusions}
\label{sec:concl}

The results presented in this paper demonstrate the importance of the
extensive R\&D and QA conducted during DIRC development and construction. 
It is apparent that without this program, the DIRC would not have
performed as well as required. 

Perhaps the most fundamental results are related to the choice of 
material:
\begin{itemize}
\item Natural fused silica materials, when exposed to a radiation
dose as low as 5-10\ krad, suffer from serious radiation damage,
resulting in substantial transmission losses in the blue and UV.
All synthetic fused silica samples, however, were found to be
sufficiently radiation hard for \babar . 
\item Some synthetic fused silica materials, made in
the form of ingots, show periodic optical inhomogeneities.
This effect can be sufficiently large to make the DIRC inoperable.
\end{itemize}

\section*{Acknowledgments}

We would like to thank H.~Kr\"{u}ger for important contributions during
his 1996 summer stay at SLAC.  We also thank R.~Reif, M.~McCulloch,
and M.~Schneider for their excellent technical help.

%% file: biblio.tex
\renewcommand{\thefootnote}{\fnsymbol{footnote}}